\documentclass[english]{article}
\usepackage[T1]{fontenc}
\usepackage[latin9]{inputenc}
\usepackage{geometry}
\usepackage{units}
\usepackage{url}
\usepackage{amsmath}
\usepackage{amssymb}
\usepackage{stmaryrd}
\usepackage{graphicx}

\makeatletter
\newcommand{\lyxaddress}[1]{
	\par {\raggedright #1
	\vspace{1.4em}
	\noindent\par}
}

\usepackage[labelfont=bf]{caption}

\makeatother

\usepackage{babel}
\begin{document}
\title{Magnetopause ripples going against the flow form azimuthally stationary
surface waves}
\author{M.~O.~Archer$^{1}${*}, M.~D.~Hartinger$^{2}$, F.~Plaschke$^{3}$,
D.~J.~Southwood$^{4}$, \& L. Rastaetter$^{5}$}
\maketitle

\lyxaddress{$^{1}$ Space and Atmospheric Physics Group, Department of Physics,
Imperial College London, London, UK. (m.archer10@imperial.ac.uk)\\
$^{2}$ Space Science Institute, Boulder, Colorado, USA. (mhartinger@spacescience.org)\\
$^{3}$ Space Research Institute, Austrian Academy of Sciences, Graz,
Austria. (ferdinand.plaschke@oeaw.ac.at)\\
$^{4}$ Space and Atmospheric Physics Group, Department of Physics,
Imperial College London, London, UK. (d.southwood@imperial.ac.uk)\\
$^{5}$ NASA Goddard Space Flight Center, Greenbelt, Maryland, USA.
(lutz.rastaetter-1@nasa.gov)}
\begin{abstract}
Surface waves process the turbulent disturbances which drive dynamics
in many space, astrophysical and laboratory plasma systems, with the
outer boundary of Earth's magnetosphere, the magnetopause, providing
an accessible environment to study them. Like waves on water, magnetopause
surface waves are thought to travel in the direction of the driving
solar wind, hence a paradigm in global magnetospheric dynamics of
tailward propagation has been well-established. Here we show through
multi-spacecraft observations, global simulations, and analytic theory
that the lowest-frequency impulsively-excited magnetopause surface
waves, with standing structure along the terrestrial magnetic field,
propagate against the flow outside the boundary. Across a wide local
time range ($09\text{\textendash}15\mathrm{h}$) the waves' Poynting
flux exactly balances the flow's advective effect, leading to no net
energy flux and thus stationary structure across the field also. Further
down the equatorial flanks, however, advection dominates hence the
waves travel downtail, seeding fluctuations at the resonant frequency
which subsequently grow in amplitude via the Kelvin-Helmholtz instability
and couple to magnetospheric body waves. This global response, contrary
to the accepted paradigm, has implications on radiation belt, ionospheric,
and auroral dynamics and potential applications to other dynamical
systems.
\end{abstract}

\section*{Introduction}

Sharp discontinuities separating regions with different physical
parameters are a key feature of space, astrophysical and laboratory
plasmas. Their dynamics about pressure-balanced equilibrium, magnetohydrodynamic
(MHD) surface waves, act as an efficient mechanism of filtering, accumulating,
and guiding the turbulent disturbances omnipresent between/through
their respective systems. Surface waves have been observed and modelled
within tokamak experiments \cite{connor98}, plasma tori surrounding
planets \cite{He2020}, the solar atmosphere (e.g. in coronal loops
\cite{li13}), the heliopause \cite{baranov92}, accretion disks \cite{stehle99},
and astrophysical/relativistic jets \cite{zhao92} to name a few.
This makes understanding surface waves of universal importance.

While many of these environments can only be remote sensed, planetary
magnetospheres (particularly that of Earth) provide the opportunity
of \textit{in situ} measurements of surface wave processes. The motion
of the outer boundary of a magnetosphere, the magnetopause, is of
primary importance in dictating global magnetospheric dynamics since
it controls the flow of mass, energy, and momentum from the solar
wind into the terrestrial system having direct and indirect space
weather impacts on the radiation belts, auroral oval, and ionosphere
\cite{summers13,elkington06,keiling16}. Surface waves on a planetary
magnetopause, which occupy the lower ends of the so-called ultra-low
frequency range (ULF; fractions of milliHertz to a few Hertz), are
excited either by pressure imbalances (typically on the dayside) or
flow shears (on the flanks) \cite{pu83,kivelson95}. Magnetopause
surface waves are thus similar to surface waves on bodies of water,
which form due to and travel in the direction of the wind \cite{phillip06,miles06}.
Since magnetopause surface waves impart momentum on the magnetosphere,
the antisunward flow of the external driver --- the shocked solar
wind --- has led to a well-accepted paradigm of the tailward propagation
of outermagnetospheric ULF waves at all local times \cite{samson71,leonovich16,plaschke16,hwang16}.
Surface waves may subsequently become non-linear via the Kelvin-Helmholtz
instability at the magnetotail (when sufficient free energy is present
to overcome magnetic tension or plasma compressibility) forming vortices
which undergo magnetic reconnection, transporting mass across the
boundary \cite{nykyri01,ma14}. The paradigm of tailward propagation
in magnetospheric dynamics is thought to hold even in response to
the rather common impulsive events that drive intense space weather
\cite{sibeck90}, for example large-scale solar wind pressure pulses
and shock waves\cite{zuo15,villante16} or smaller ($\mathrm{R_{E}}$
scale or less) kinetically-generated bow shock phenomena like magnetosheath
jets \cite{plaschke18}. The models predict an exception, in agreement
with several observations, only at the early post-noon magnetopause,
since pressure fronts aligned with the Parker spiral interplanetary
magnetic field (IMF) strike this region before the pre-noon sector.
Reported instances of sunward propagating ULF waves have been attributed
to internal processes, such as energetic particle instabilities \cite{Constantinescu09}
or changes in the magnetotail configuration \cite{nielsen84,eriksson08}.

In physics, a common approach to understanding a dynamical system
is to determine its normal modes. These form in a magnetosphere when
system-scale MHD waves become trapped through reflection by boundaries
or turning points. Transverse Alfv\'{e}n waves, propagating along
field lines due to magnetic tension forces, are reflected by the highly
conductive ionosphere resulting in standing waves akin to those on
a guitar string \cite{southwood74}. Fast magnetosonic waves, driven
by correlated thermal and magnetic pressure gradients, trapped radially
form so-called cavity/waveguide modes \cite{kivelson84,kivelson85},
somewhat similar to the resonances of wind instruments. These magnetospheric
normal modes due to MHD body waves have been well studied and tend
to conform to the aforementioned paradigm \cite{samson71,leonovich16,plaschke16,hwang16}.
However, it had long been proposed that magnetopause surface waves
propagating along the terrestrial magnetic field in response to impulsive
pressure variations might too reflect at the northern and southern
ionospheres, forming a magnetopause surface eigenmode (MSE) somewhat
analagous to the vibrations of a drum's membrane \cite{chen74}. The
theory of these standing waves has been developed using ideal incompressible
MHD in box model magnetospheres \cite{plaschke11}. Despite their
simplicity, these models have been able to reproduce many features
captured by more advanced global MHD simulations \cite{hartinger15}.
For example, MSE frequencies near the subsolar magnetopause can be
approximated in the limit $k_{\phi}\ll k_{\parallel}$ as (equation~6
of \cite{archer15}, using pressure balance at the magnetopause)

\begin{align}
\omega & \approx k_{\parallel}\frac{B_{sph}}{\sqrt{\mu_{0}\rho_{msh}}}\label{eq:mse-frequency}\\
 & \approx k_{\parallel}\sqrt{2\frac{\rho_{sw}}{\rho_{msh}}}v_{sw}\label{eq:mse-frequency-sw}
\end{align}
for angular frequency $\omega$, wavenumber $k$, magnetic field strength
$B$, mass density $\rho$ and speed $v$ with subscripts $sw$, $msh$,
and $sph$ corresponding to the solar wind, magnetosheath, and magnetosphere
respectively. MSE thus constitute the lowest frequency normal mode
of the magnetospheric system, given the smaller phase speeds and wavenumbers
than other modes. Indeed, equation~\ref{eq:mse-frequency-sw} yields
fundamental frequencies below $2\,\mathrm{mHz}$ and thus evanescent
scales that highly penetrate the dayside magnetosphere \cite{archer15,hartinger15}.
However, MSE are thought to be strongly damped due to the finite thickness
of the boundary, perhaps persisting for only a few wave periods \cite{chen74,hartinger15,kozyreva19}.
Direct evidence of MSE was discovered only recently \cite{archer19},
likely due to the observational challenges in unambiguously demonstrating
such a low frequency normal mode has been excited. Fortuitous multi-spacecraft
observations of the response to an isolated, broadband magnetosheath
jet revealed narrowband magnetopause oscillations and magnetospheric
ULF waves that were in excellent agreement with the theoretical predictions
of MSE and could not be explained by other known mechanisms. These
observations in the mid--late morning sector strikingly showed no
azimuthal motion of the boundary despite the expectation that surface
waves be advected tailward \cite{pu83,sibeck90}, hinting that this
eigenmode may challenge the usual paradigm. It is currently unclear
how to reconcile this with current models, especially since (unlike
meridionally) there is no boundary azimuthally for surface waves to
reflect against to reverse course.

In this paper we address this conundrum by considering surface waves'
energy flux through spacecraft observations, global MHD simulations,
and analytic MHD theory in order to explain the resonant response
of the magnetospheric system globally. We show that magnetopause surface
waves propagate against the flow forming an azimuthally stationary
wave across a wide local time range.

\section*{Results}

\subsection*{Spacecraft observations}

\begin{figure*}
\begin{centering}
\noindent \makebox[\textwidth]{\includegraphics{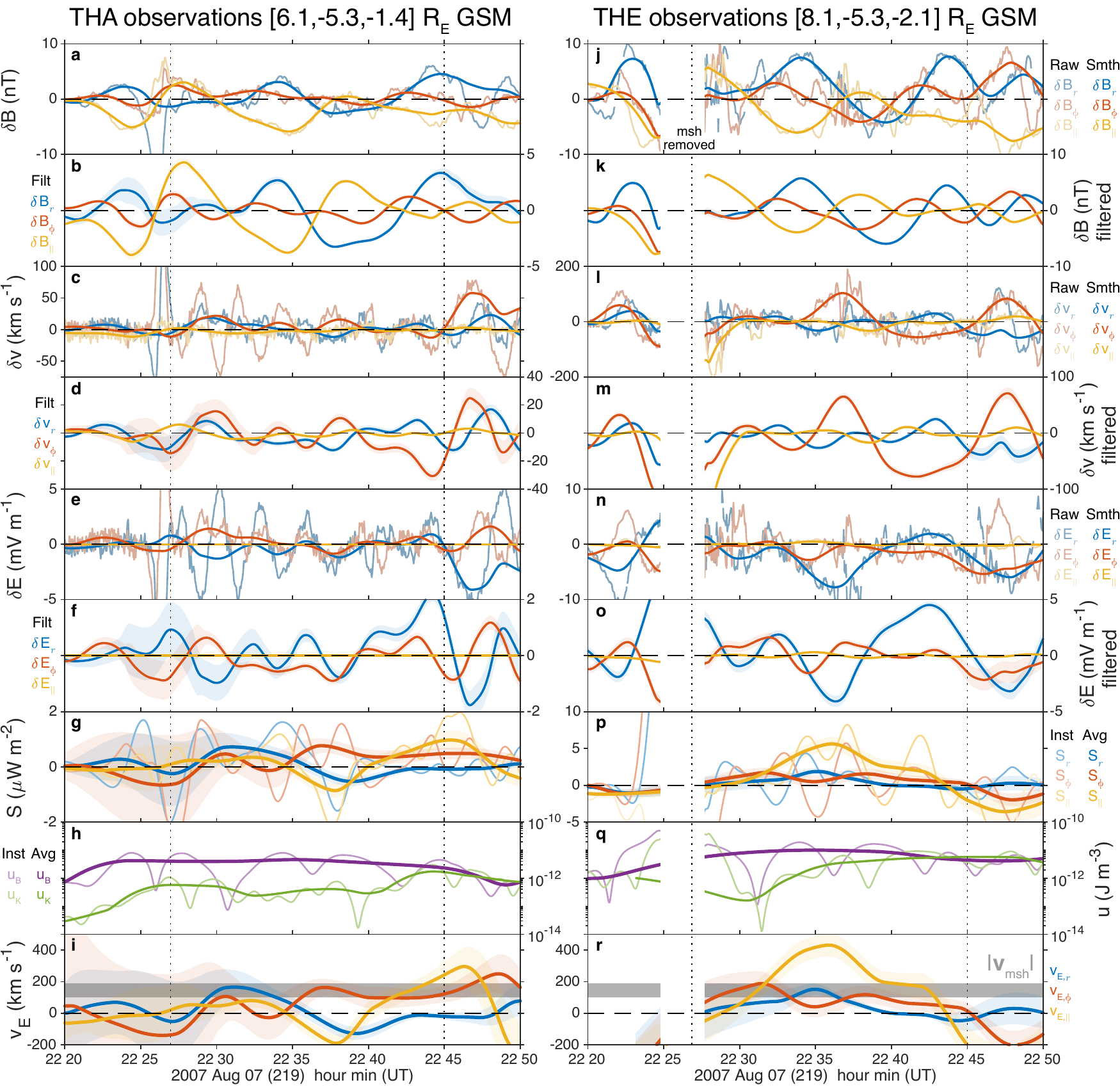}}
\par\end{centering}
\caption{\textbf{THEMIS spacecraft time series observations in the magnetosphere}.
Shown for THA (panels a--i) and THE (panels j--r). From top to bottom
the first set of panels show perturbations in the magnetic (a--b,
j--k), ion velocity (c--d, l--m), and electric (e--f, n--o) fields.
In these vertical pairs, top panels show the raw (thin) and LOESS
smoothed (thick) data, whereas the bottom panels show the latter once
detrended. Subsequent panels depict the Poynting vector (g, p) and
energy density (h, q), showing instantaneous (thin) and time-averaged
(thick) values. Finally the energy velocity (i, r) is shown compared
to the absolute magnetosheath flow speed at THB (grey). Throughout,
standard errors are indicated by shaded areas. Vertical dotted lines
demark the times of little upstream pressure variations following
the isolated impulsive jet that triggered this event. \label{fig:spacecraft-timeseries}}
\end{figure*}

\begin{figure}
\begin{centering}
\noindent \makebox[\textwidth]{\includegraphics{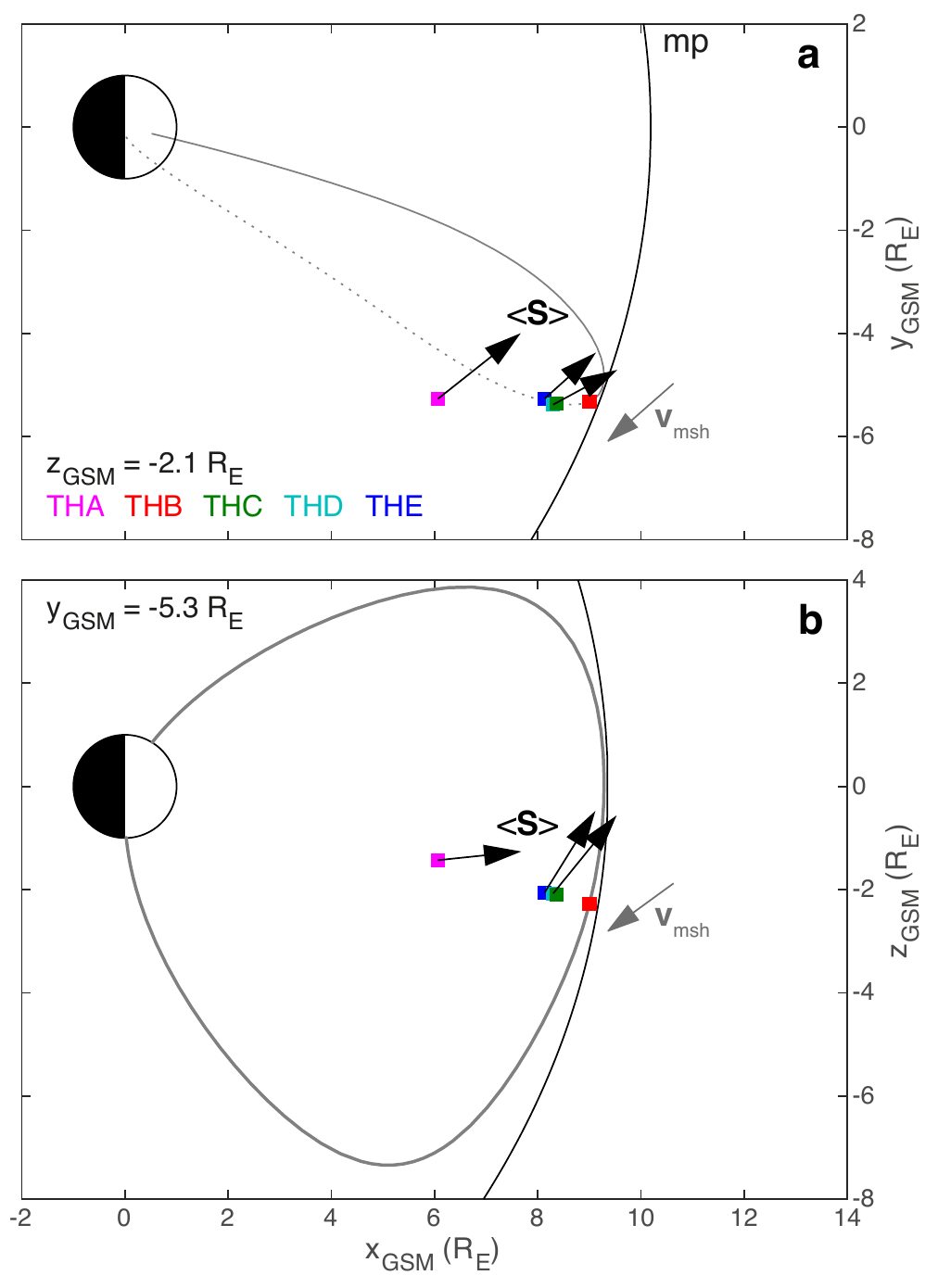}}
\par\end{centering}
\caption{\textbf{Directions of the time-averaged Poynting vectors at each THEMIS
spacecraft location}. These are displayed in the $z_{GSM}=-2.1\,\mathrm{R_{E}}$
(a) and $y_{GSM}=-5.3\,\mathrm{R_{E}}$ (b) planes. Coloured squares
represent the THEMIS spacecraft positions. Black arrows emanating
from these indicate normalised Poynting vectors. Grey arrows depict
the magnetosheath flow velocity direction. A representative geomagnetic
field line (grey) and model magnetopause location (black) are also
shown.\label{fig:spacecraft-locations}}
\end{figure}

We use Time History of Events and Macroscale Interactions (THEMIS)
\cite{angelopoulos08} spacecraft observations from satellites A--E
(THA--THE) during the previously reported event of MSE triggered
by a magnetosheath jet on 07~August~2007. See the spacecraft observations
section of Methods for further details of instruments and techniques
employed. The spacecraft were located at $\sim$09:30~MLT (magnetic
local time) in a string-of-pearls formation. For context, at 22:25~UT
an isolated $\sim100\,\mathrm{s}$ magnetosheath jet occurred upstream
of the magnetopause which was followed by a period of $\sim18\,\mathrm{min}$
with little pressure variations (demarked by vertical dotted lines)
until another jet occurred at 22:45~UT (Figure~2d of \cite{archer19}).
The magnetopause moved in response to the jet, undergoing two boundary
oscillations at $1.8\,\mathrm{mHz}$ corresponding to the fundamental
mode MSE (Figures 2g and 3b of \cite{archer19}). Figure~\ref{fig:spacecraft-timeseries}
shows magnetospheric observations by the THA (panels a--i) and THE
(panels j--r) spacecraft of the magnetic (panels~a, j), velocity
(panels~c, l), and electric fields (panels~e, n). Dynamic spectra
of these using the continuous wavelet transform can also be found
in Supplementary Figure~1 revealing the $1.8\,\mathrm{mHz}$ fundamental
mode MSE (clearest in the compressional magnetic field components
at both spacecraft) and $3.3\,\mathrm{mHz}$ second harmonic MSE (e.g.
in the perpendicular components of the magnetic field), as well as
a $6.7\,\mathrm{mHz}$ fundamental toroidal standing Alfv\'{e}n wave
at THA (azimuthal velocity / radial electric field) \cite{archer19}.
THD spacecraft observations proved similar to THE, and the other spacecraft
encountered the magnetosheath too often for use here. We aim to measure
the Poynting vector and corresponding energy velocity associated with
MSE, concepts which are further explained in the Poynting's theorem
for MHD waves section of Methods. This necessitates extracting the
associated wave perturbations from the data, removing noise and other
signals as described in the time-based filtering section of Methods,
resulting in the filtered magnetic (panels~b, k), velocity (panels~d,
m), and electric fields (panels~f, o) shown in Figure~\ref{fig:spacecraft-timeseries}.
These are then used to determine energy densities and fluxes. Between
the two dotted lines (which indicate the times of little upstream
pressure variations) all spacecraft observed time-averaged Poynting
vectors with components consistently azimuthally eastward and a slight
tendency towards radially outwards too (panels~g, p). This was also
evident at the MSE frequencies in the Poynting vectors computed using
the wavelet transforms (see fourier and wavelet techniques of Methods
for details) which are shown in Supplementary Figure~1 in time-frequency
and in Supplementary Figure~2 as a function of frequency by averaging
over the interval. The average Poynting directions at each spacecraft
location are shown in Figure~\ref{fig:spacecraft-locations}, showing
excellent agreement across all spacecraft in the equatorial plane
(panel~a). These observations show MSE do not conform to the typical
ULF wave paradigm of tailward propagation \cite{sibeck90,keiling16}
(waveguide modes' Poynting fluxes are directed tailward or have no
net azimuthal component \cite{elsden15,elsden19,wright20}; Kelvin-Helmholtz
generated surface waves travel tailward and radiate energy into the
magnetosphere \cite{juninger85,sakurai01}). The wave energy density
(Figure~\ref{fig:spacecraft-timeseries}~h, q) is dominated by the
magnetic component, though the kinetic energy becomes comparable later
in the interval. The waves' azimuthal energy velocity (Figure~\ref{fig:spacecraft-timeseries}~i,
r) is comparable to the flow speed in the magnetosheath (absolute
value in grey) but oppositely directed, as indicated in Figure~\ref{fig:spacecraft-locations}a,
suggesting the two forms of opposing energy flux might balance one
another out and result in no net energy flow, i.e. an azimuthally
stationary wave. This potentially may be behind the lack of azimuthal
propagation in the observed boundary motion during this interval \cite{archer19}
and may hold the key to how MSE are even possible away from the noon
meridian.


Since surface waves are formulated as collective magnetosonic waves
on both sides of the boundary, the component of the Poynting vector
towards the magnetopause might be understood in terms of the magnetosonic
dispersion relation (equation 7 of \cite{pu83})
\begin{equation}
k_{r}^{2}=-k_{\phi}^{2}-k_{\parallel}^{2}+\frac{\omega^{4}}{\omega^{2}v_{A}^{2}+c_{s}^{2}\left(\omega^{2}-\left[\mathbf{k}\cdot\mathbf{v}_{A}\right]^{2}\right)}\label{eq:dispersion}
\end{equation}
, where $\mathbf{v}_{A}$ is the Alfv\'{e}n velocity and $c_{s}$
the speed of sound. Under incompressibility the last term is neglibigle
resulting in a purely imaginary $k_{r}$ and thus evanescence over
similar scales to the length of the geomagnetic field lines. We relax
this assumption and use a complex frequency $\omega=\omega_{\mathfrak{Re}}-i\gamma$
with damping rate $\gamma>0$, since surface waves on a boundary of
finite thickness are thought to be damped \cite{chen74}. For the
magnetospheric side, the phase of the last term in equation~\ref{eq:dispersion}
is negative (approximately twice that of $\omega$ as the plasma beta
is small) and thus $k_{r}^{2}$ has a negative imaginary component.
This implies, for a physically reasonable solution with zero amplitude
at infinity, $k_{r}$ should have a small real component pointed towards
the magnetopause. We estimate that a damping ratio $\gamma/\omega_{\mathfrak{Re}}=0.15$
should result in radial phase velocity components of $\sim10\,\mathrm{km}\,\mathrm{s}^{-1}$
(and between $1\text{\textendash}60\,\mathrm{km}\,\mathrm{s}^{-1}$
for $\gamma/\omega_{\mathfrak{Re}}=0.02\text{\textendash}1$ \cite{hartinger15,kozyreva19}),
i.e. considerably smaller than the Alfv\'{e}n speed of $\sim1000\,\mathrm{km}\,\mathrm{s}^{-1}$
and consistent with the average observed radial energy velocities
of $9\text{\textendash}46\,\mathrm{km}\,\mathrm{s}^{-1}$. This sense
of propagation is opposite to what is expected for a Kelvin-Helmholtz
unstable boundary, where the sign of $\gamma$ is reversed (being
a growth rate) and thus results in energy radiating into the magnetosphere.
By conservation of energy flux across the boundary, the Poynting vector
component towards the boundary would imply that damped magnetopause
surface waves lose some of their energy to the magnetosheath. This
energy pathway would be in addition to the theorised irreversible
conversion of surface wave energy to the Alfv\'{e}n continuum \cite{kozyreva19}.

THE and THD both observed significant field-aligned energy flux also,
seemingly less prominent at THA. One might naively expect no field-aligned
energy flux for a standing surface wave. However, considering this
is a dynamical mode involving surface waves reflecting and interfering
along the field under asymmetric conditions and driving, a resultant
flux in this direction may be expected. For example, reflection at
the ionospheres is not perfect nor is the absorption north-south symmetric
\cite{southwood00}. This will yield a superposition of standing and
propagating waves with a ``null point'', shifted slightly from the
standing wave's nodes/antinodes, either side of which some resultant
wave energy propagates to the respective ionospheres where it is dissipated
\cite{allan82}. The field-aligned flux at THE and THD peaks approximately
one MSE bounce time after the driving jet. The corresponding energy
velocity is consistent with the surface wave phase speed (equation~\ref{eq:mse-frequency}).
It therefore seems plausible that these signatures are due to both
the dipole tilt (Figure~\ref{fig:spacecraft-locations}b) resulting
in different reflectances in both hemispheres and the localised driver
exciting multiple harmonics with different relative phases causing
shifts in the interference pattern. To the first point, the dipole
tilt for this event was $17^{\circ}$, hence different conductances
in the northern and southern ionospheres would be expected. Further,
THA's footpoint ($66^{\circ}$ geomagnetic latitude) could also have
a different conductance to that for THD and THE ($71^{\circ}$, i.e.
near/in the auroral oval) \cite{ridley04}, which could result in
a difference in the proportion of wave energy reflected back to the
spacecrafts' respective locations. To the second point, the wavelet
transform demonstrates differences in the field-aligned Poynting fluxes
for the two harmonics, most clearly shown in Supplementary Figure~2
where averaging over time has been applied. The time-averaged Poynting
flux at the fundamental MSE frequency of $1.8\,\mathrm{mHz}$ has
a component in the direction of the geomagnetic field at all spacecraft,
indicating the spacecraft were all located above this harmonic's ``null
point''. The direction of the Poynting vectors agreed to within $26^{\circ}$
and thus are consistent taking noise into account. However, at the
second harmonic MSE of $3.3\,\mathrm{mHz}$, while the Poynting vectors'
projections in the equatorial plane are similar (to within $9^{\circ}$),
along the field we find that THE/THD observed southward fluxes whereas
at THA they were slightly northward (though not statistically significant
from noise). A second harmonic wave has a node in displacement near
the equator, thus at this frequency the spacecraft observations are
sensitive to which side of the ``null point'' they are located.
In addition, smaller wavelength surface waves are less penetrating
into the magnetosphere (equation~\ref{eq:dispersion}) which would
weaken the signal at THA's location. We conclude that THA was very
close to the second harmonic MSE's ``null point'' whereas THE/THD
were slightly below it. Nonetheless, the main result of interest in
this paper, i.e. the fluxes radially and azimuthally, are in good
agreement across all spacecraft at both frequencies. MSE's field-aligned
energy flux may have implications on energy deposition in the ionosphere
and warrants investigation in future work.

The above analysis was limited to a relatively short interval of confirmed
MSE activity following an isolated magnetosheath jet. However, several
other jets were also observed on this day and it was noted that similarly
directed Poynting vectors followed many of them. We therefore take
a wider interval and compute the time-averaged Poynting vector as
a function of frequency, as detailed in the fourier and wavelet techniques
section of Methods. This was performed for THA as it was the only
spacecraft to experience uninterupted magnetospheric observations.
Supplementary~Figure~3 shows that at MSE frequencies (dotted lines)
the radial and azimuthal Poynting vector components were statistically
significant and in agreement with the previous results, namely outward
and eastward. The parallel Poynting flux is positive at MSE frequencies
indicating that THA was overall located above the respective ``null
points'' of these waves \cite{allan82}. We note that there is a
reversal of the parallel Poynting flux around the local Alfv\'{e}n
mode frequency ($6.7\,\mathrm{mHz}$) thus unrelated to MSE. The typical
tailward energy flow paradigm emerges only at much higher frequencies
($>10\,\mathrm{mHz}$).

\subsection*{Global MHD simulations}

\begin{figure*}[p]
\begin{centering}
\noindent \makebox[\textwidth]{\raisebox{0pt}[23cm]{\includegraphics{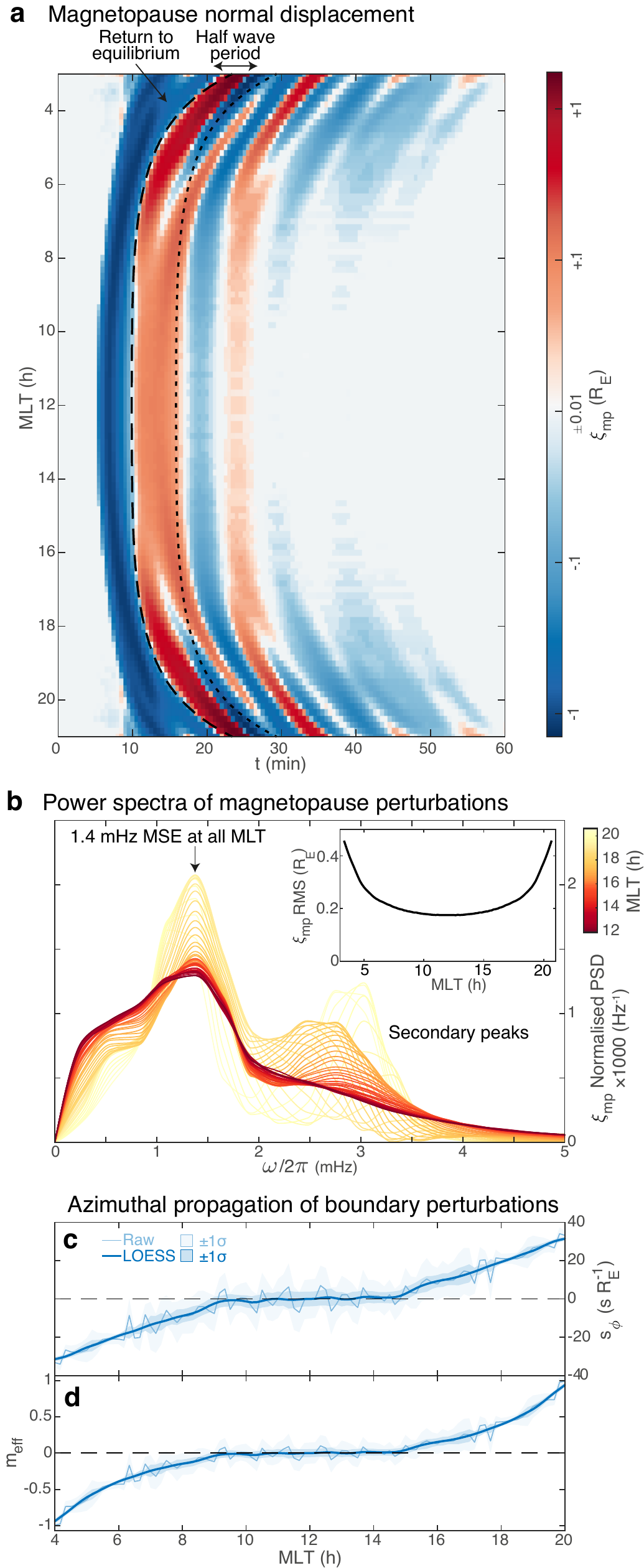}}}
\par\end{centering}
\caption{\textbf{Magnetopause motion in MHD simulation. a) Normal displacement
with magnetic local time (MLT)}. b) Spectra of the displacement normalised
by total power for each MLT (colour scale), with the root mean squared
(RMS) also given inset. Azimuthal slowness (c) and effective wavenumber
(d) of filtered magnetopause perturbations for the raw data (lighter)
as well as after applying MLT smoothing (darker), with corresponding
standard errors shown in both cases as shaded areas.\label{fig:Magnetopause-motion}}
\end{figure*}

To further test our hypotheses from the THEMIS observations, global
MHD simulations of the global magnetospheric response to a $1\,\mathrm{min}$
large-scale solar wind density pulse are now employed (see Global
MHD simulations in Methods). This reproduces a previous simulation
\cite{hartinger15}, where the subsolar response could only be explained
by MSE and not other mechanisms. The normal displacement of the magnetopause
in the XY plane is shown in Figure~\ref{fig:Magnetopause-motion}a.
This highlights the dayside magnetopause undergoes a strong compression
when the pulse arrives, rebounds returning to equilibrium (dashed
line) but overshoots, subsequently undergoing damped oscillations.
Results are identical on both flanks due to the symmetry of the system
and driver. The oscillations' primary frequency is $1.4\,\mathrm{mHz}$
at all local times (panel~b), consistent with a fundamental MSE \cite{hartinger15}.
A secondary peak in the spectra, not previously reported, grows further
downtail between $2.5\text{\textendash}3.3\,\mathrm{mHz}$. Both spectral
peaks are associated with the damped oscillations and not the broadband
initial compression/rebound motions, as checked by a wavelet transform.
The secondary mode is likely due to (and at the frequency which maximises)
the Kelvin-Helmholtz instability given the increasing flow shear across
the boundary down the flanks \cite{merkin13}. Both modes become larger
in amplitude (panel~b and inset) and persist longer (panel~a) further
downtail, though the primary mode is always dominant. This suggests
that MSE at $1.4\,\mathrm{mHz}$, which originates on the dayside
magnetopause, seeds fluctuations which subsequently grow via Kelvin-Helmholtz
in the flanks despite not being at the instability's peak growth frequency.

We now investigate the propagation of the $1.4\,\mathrm{mHz}$ magnetopause
oscillations, extracted as described in the time-based filtering section
of Methods. From Figure~\ref{fig:Magnetopause-motion}a it appears
that across much of the dayside the waves do not propagate azimuthally
(the phase fronts are vertical), whereas it is clear in the flanks
that tailward propagating waves are present (inclined fronts, see
also Supplementary Movie~1). Here we quantify this propagation via
the azimuthal slowness $s_{\phi}$ (reciprocal of apparent phase speed,
see slowness in Methods) since the slowness vector is always normal
to phase fronts \cite{gaiser90}. The results are shown in Figure~\ref{fig:Magnetopause-motion}c.
This reveals that between $\sim$09--15h~MLT the slowness is zero
and thus the surface wave is apparently azimuthally stationary in
this region. Further down both flanks though the usual tailward motion
is recovered. It may be instructive to express effective local azimuthal
wavenumbers $m_{eff}=s_{\phi}\omega r_{mp}\left(\phi\right)$, shown
in panel~d ($r_{mp}\left(\phi\right)$ is the magnetopause geocentric
distance at each azimuth). Care must be taken in interpreting these
since the magnetopause crosses L-shells and is not azimuthally symmetric,
so the dependence cannot be expressed simply as $\exp\left(im\phi\right)$
everywhere. Instead a superposition of wavenumbers will be present,
with $m_{eff}$ capturing the local azimuthal propagation of the overall
phase \cite{degeling14}. $\left|m_{eff}\right|$ is zero in the stationary
wave region, rises slowly to $\sim0.5$ by the terminator, then more
rapidly increases to $\sim1$ within a further 2~h of LT. This global
structure cannot be attributed to the driver, since the intersection
of the pressure pulse with the magnetopause on arrival spans 08--16h~MLT,
i.e. larger than the stationary region.

\begin{figure}
\begin{centering}
\noindent \makebox[\textwidth]{\includegraphics{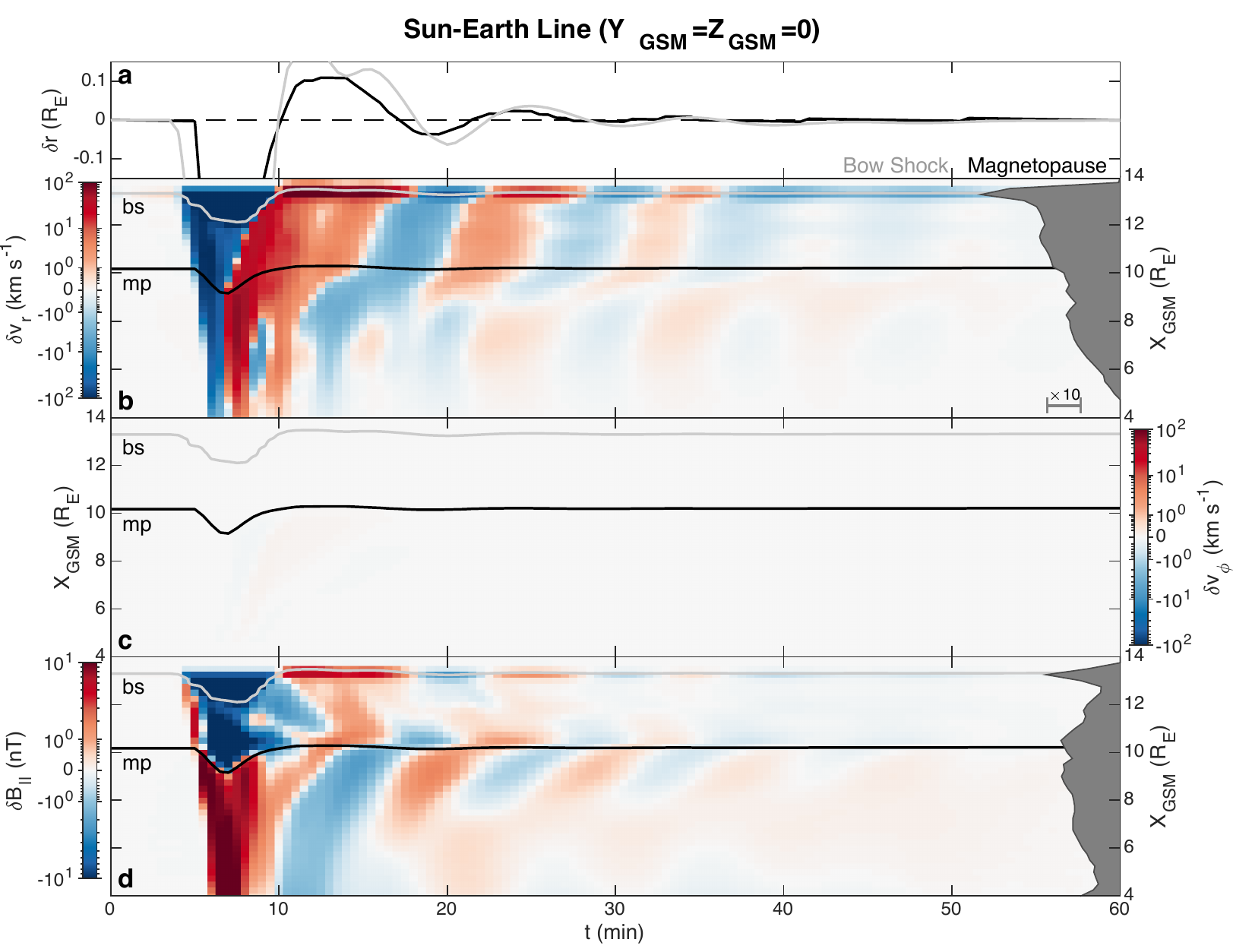}}
\par\end{centering}
\caption{\textbf{Unfiltered perturbations in MHD simulation along the Sun-Earth
line}. Panel~a shows motion of the bow shock (grey) and magnetopause
(black) about equilibrium. Subsequent panels show perturbations in
the b) radial velocity, c) azimuthal velocity, and d) compressional
magnetic field components (note the bi-symmetric log scale). Median
absolute perturbations by distance are displayed to the right as the
dark grey areas on a logarithmic scale. The bow shock (light grey)
and magnetopause (black) locations are also plotted.\label{fig:subsolar}}
\end{figure}

\begin{figure}
\begin{centering}
\noindent \makebox[\textwidth]{\includegraphics{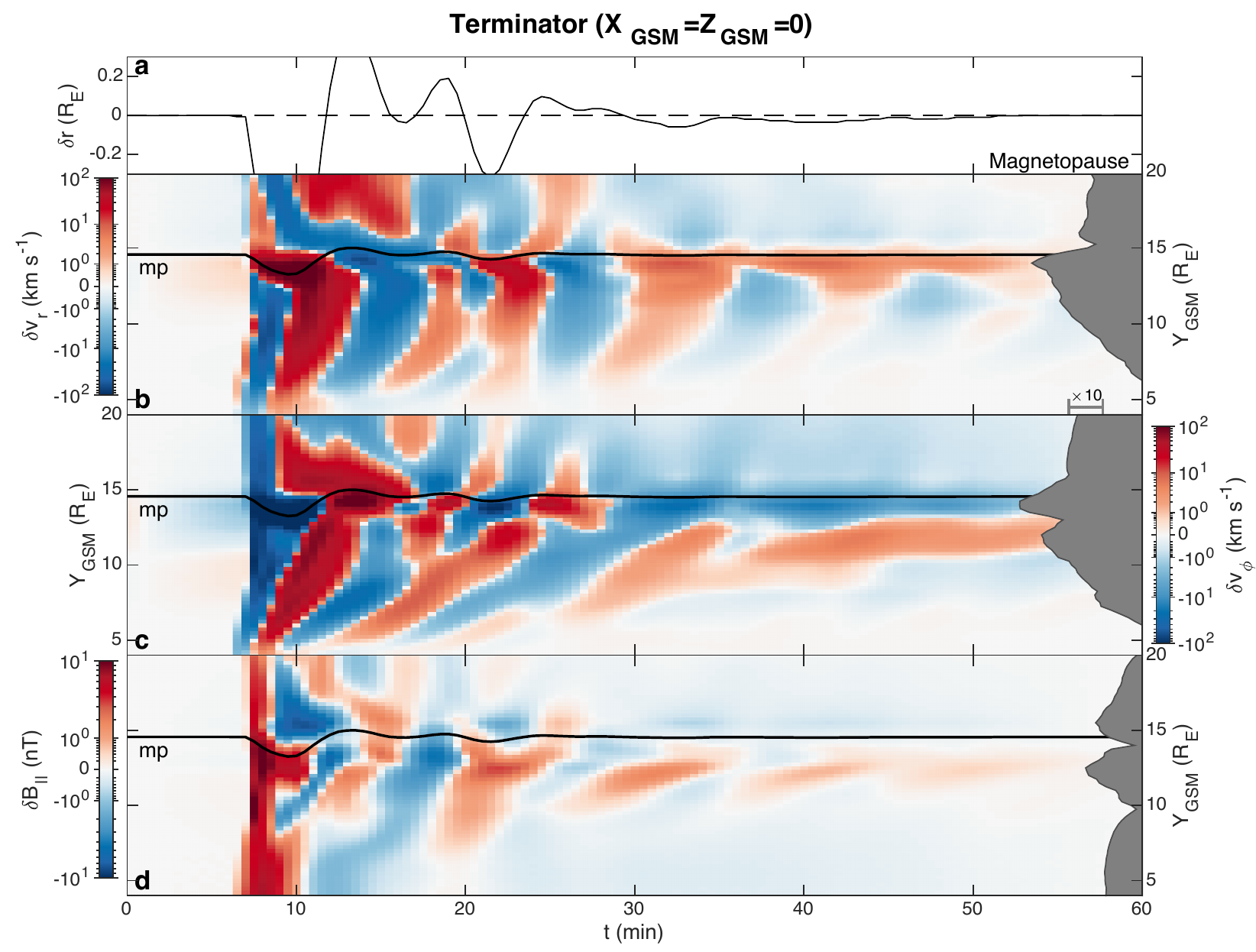}}
\par\end{centering}
\caption{\textbf{Unfiltered perturbations in MHD simulation along the equatorial
terminator}. Formatting is the same as Figure~\ref{fig:subsolar}.
Panel~a shows motion of the magnetopause (black) about equilibrium.
Subsequent panels show perturbations in the b) radial velocity, c)
azimuthal velocity, and d) compressional magnetic field components
(note the bi-symmetric log scale). Median absolute perturbations by
distance are displayed to the right as the dark grey areas on a logarithmic
scale. The magnetopause (black) location is also plotted.\label{fig:terminator}}
\end{figure}

We now look at the grid point data within the simulation. Supplementary
Movie~1 shows the compressional magnetic field perturbations in the
XY and XZ planes. Figure~\ref{fig:subsolar} shows boundary (panel~a),
radial (b) and azimuthal (c) velocity, and compressional magnetic
field (d) perturbations along the Sun-Earth line These demonstrate
that the arrival of the pressure pulse and inward magnetopause motion
launches a compressional wave into the dayside magnetosphere which
reflects at/near the simulation's inner boundary and subsequently
leaks into the magnetosheath where it dissipates. This all happens
within $\sim2\,\mathrm{min}$ (i.e. before the magnetopause has finished
rebounding) in agreement with the magnetosonic speed profile. Such
a short timescale provides further evidence (in addition to that in
\cite{hartinger15}) that the subsequent magnetopause oscillations
on the dayside cannot be attributed to cavity/waveguide modes as the
lowest frequency (quarter wavelength \cite{mann99}) mode should be
$\gtrsim4\,\mathrm{mHz}$. Azimuthal velocities are negligible, hence
there is no evidence of toroidal Alfv\'{e}n waves in this region.
The MSE signatures instead are radial plasma motions and associated
compressions/rarefactions of the magnetic field, both of which decay
in amplitude with distance from the magnetopause as indicated by the
median absolute perturbations (grey areas in Figure~\ref{fig:subsolar}).
The phase fronts, however, are not purely evanescent and can be seen
propagating towards the magnetopause on the magnetospheric side. This
occurs rather slowly though at around $30\text{\textendash}40\,\mathrm{km}\,\mathrm{s}^{-1}$
near the boundary, in agreement with the estimates due to damping
made earlier. Deeper into the magnetosphere more evanescent and less
propagating behaviour is found, as expected from equation~\ref{eq:dispersion}
due to the greater Alfv\'{e}n speeds. The magnetic field perturbations
on either side of the boundary are in approximate antiphase with one
another throughout the dayside (note the magnetopause thickness in
the simulation is $\sim1.5\,\mathrm{R_{E}}$, considerably larger
than in reality since gyroradius-scales are not resolved \cite{berchem82}).
There is evidence of large-scale bow shock motion related to MSE,
a consequence which has not been considered before. At the subsolar
point (see Figure~\ref{fig:subsolar}) the bow shock lags the magnetopause
motion by $\sim1\,\mathrm{min}$, consistent with the fast magnetosonic
travel time through the magnetosheath, confirming that the resonance
is occurring at the magnetopause and subsequently driving the shock
oscillations. This lag occurs because magnetosheath plasma is highly
compressible \cite{plaschke11,archer15}, thereby deviating from the
evanescent behaviour expected under incompressibility. Since both
the magnetopause and bow shock move asynchronously, the patterns present
in the subsolar magnetosheath are somewhat complicated. These could
be explored further in the future.

In Supplementary Movie~1, magnetic field perturbations in the equatorial
plane are in phase across much of the dayside showing little evidence
of azimuthal propagation, confirming that $k_{\phi}$ is much smaller
than $k_{r}$ and $k_{\parallel}$. Tailward travelling disturbances
can be seen emanating from the oscillations near the dayside magnetopause
only at $\sim$09h and 15h~MLT, hence are associated with the propagating
surface waves discussed earlier. Supplementary Movie~2 separates
out these two regimes for further clarity. Further down the flanks,
at $\sim$07h and 17h~MLT, structure normal to the magnetopause emerges
with strong peaks/troughs $\sim2R_{E}$ inwards from the boundary.
Figure~\ref{fig:terminator} shows cuts along the terminator. This
reveals, in addition to the surface waves, the presence of a quarter
wavelength waveguide mode \cite{mann99,wright20} (at the magnetopause
there is a $\delta v_{r}$ antinode and $\delta B_{\parallel}$ node;
$\delta B_{\parallel}$ exhibits nodal structure radially). The waveguide
mode couples to a toroidal Alfv\'{e}n mode \cite{samson92} at $Y_{GSM}\sim11.5\,\mathrm{R_{E}}$
($\delta v_{\phi}$ antinode). These two modes occur at the same $\sim10\,\mathrm{min}$
period as the surface waves that originate at the subsolar point.
Therefore, MSE can couple to body eigenmodes in regions of the magnetosphere
where their frequencies sufficiently match (checked through time-of-flight
estimates). Magnetospheric Alfv\'{e}n speed profiles are highly variable
and significantly alter the eigenfrequencies of both body modes \cite{archer15b},
thus we expect that whether and where this coupling may occur will
vary substantially.

In the XZ plane, Supplementary Movie~1 reveals that the $\sim10\,\mathrm{min}$
period oscillations do not extend beyond the cusps into the northern
and southern tail lobes. The waves are thus confined to closed magnetic
field lines, further backing the surface eigenmode interpretation.

\begin{figure*}
\begin{centering}
\noindent \makebox[\textwidth]{\includegraphics{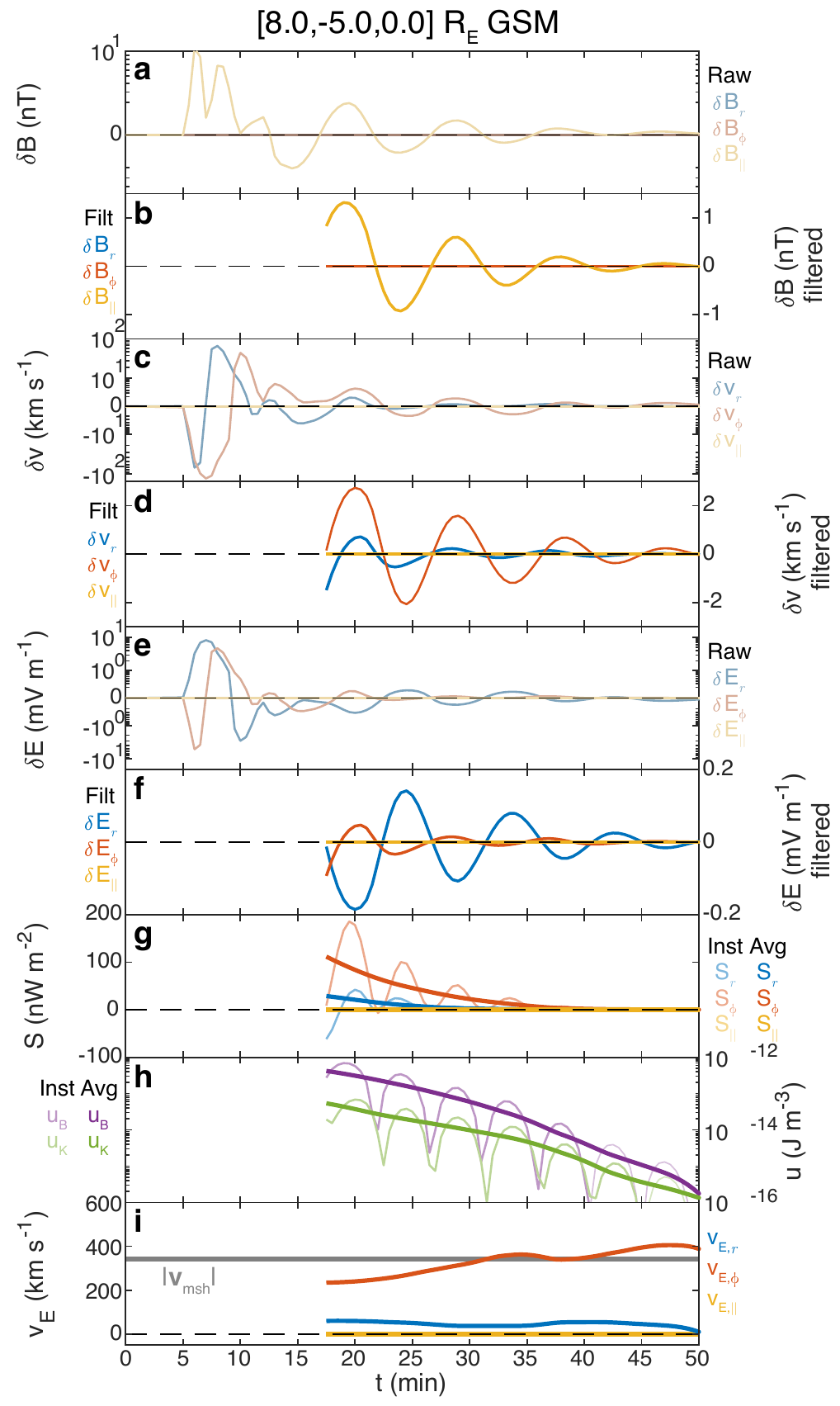}}
\par\end{centering}
\caption{\textbf{Virtual spacecraft observations within MHD simulation}. Displayed
in a similar format to Figure~\ref{fig:spacecraft-timeseries}. From
top to bottom the first set of panels show perturbations in the magnetic
(a--b), ion velocity (c--d), and electric (e--f) fields. In these
vertical pairs, top panels show the raw data, whereas the bottom panels
show the filtered data. Subsequent panels depict the Poynting vector
(g) and energy density (h), showing instantaneous (thin) and time-averaged
(thick) values. Finally the energy velocity (i) is shown compared
to the absolute magnetosheath flow speed (grey). Note the bi-symmetric
log scale on panels a, c, and e.\label{fig:virtual-spacecraft}}
\end{figure*}

\begin{figure}
\centering{}\noindent \makebox[\textwidth]{\includegraphics{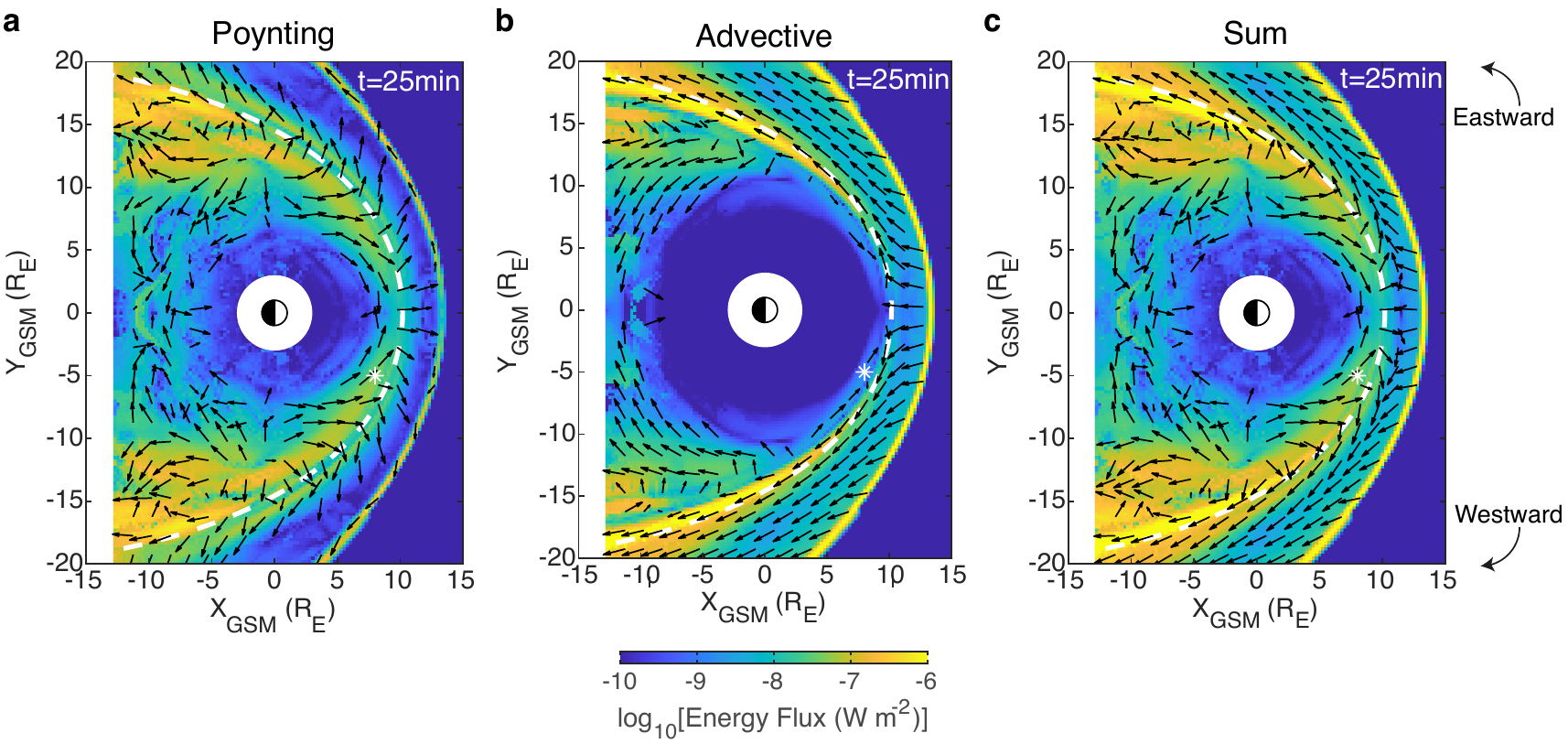}}\caption{\textbf{Wave energy flux maps}. Panels show time-averaged wave Poynting
(a), advective (b), and total (c) energy fluxes. Magnitude (colour)
and direction (arrows) are shown along with the equilibrium magnetopause
(white dashed) and virtual spacecraft location from Figure~\ref{fig:virtual-spacecraft}
(star).\label{fig:maps}}
\end{figure}

\begin{figure}
\centering{}\noindent \makebox[\textwidth]{\includegraphics{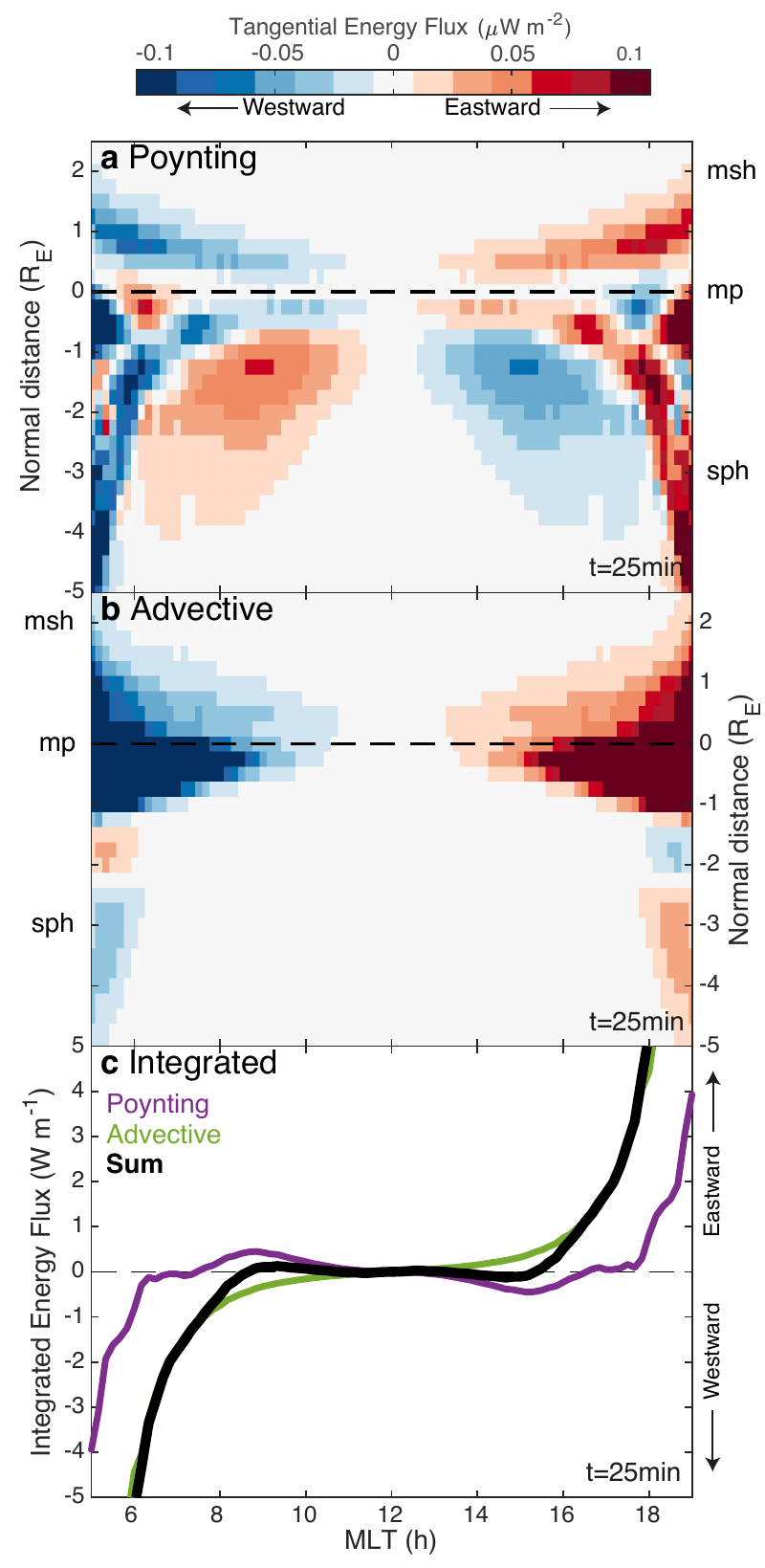}}\caption{\textbf{Wave energy fluxes tangential to the magnetopause}. Panels
show Poynting (a) and advective (b) energy fluxes tangential to the
magnetopause along magnetopause normals. Integrals along the normal
are shown in panel c for the Poynting (purple) and advective (green)
fluxes along with their sum (black).\label{fig:tangential}}
\end{figure}

We finally study energy propagation throughout the simulation. Figure~\ref{fig:virtual-spacecraft}
shows results from a virtual spacecraft in the magnetosphere close
to the boundary at roughly the same location as THE. Computing the
Poynting vector as before (panel~g) shows it to be directed azimuthally
towards the subsolar point and slighly radially outwards, similarly
to the observations. This corresponds to an energy velocity (panel~i)
approximately equal but opposite to the background magnetosheath flow
speeds (grey) at this local time, like in the observations.

Figure~\ref{fig:maps} shows equatorial maps of the time-averaged
Poynting (panel~a) and advective (panel~b) energy fluxes as well
as the sum of the two (panel~c). Within the magnetosphere, the Poynting
vectors are directed azimuthally towards the subsolar point across
the entire dayside, flipping direction at around the terminator to
recover the more usual tailward energy flux associated with Kelvin-Helmholtz
generated surface waves and waveguide modes \cite{juninger85,sakurai01,elsden15}.
Later within the simulation, however, this point of reversal does
slowly move slightly towards noon by $\sim$1h of MLT on both flanks
as the wave energy dissipates. A component directed towards the magnetopause
is also present across the dayside until well into the flanks and
the continuity of this energy flux into the magnetosheath is apparent.
The advective energy flux in Figure~\ref{fig:maps}b consists predominantly
of the tailward flow throughout the magnetosheath. Therefore, the
sum of the two clearly shows across the dayside that energy fluxes
tangential to the magnetopause are in opposition to one another on
either side. A small amount of energy flows across the boundary from
the magnetosphere into the magnetosheath, which will then be swept
downtail.

We therefore investigate the potential balance of tangential energy
fluxes on either side of the magnetopause in Figure~\ref{fig:tangential}.
At each local time we construct rays normal to the equilibrium magnetopause
and interpolate the time-averaged Poynting (panel~a) and advective
(panel~b) energy fluxes, taking the component tangential to the boundary.
Integrating these along the normal, we arrive at panel~c showing
the total tangential energy flux across both sides of the magnetopause.
This demonstrates the tailward energy flow due to advection (green)
and opposing Poynting flux (purple) across the dayside. Taking the
sum of these shows that they cancel out between 08:40--15:20~MLT,
i.e. the local time range for which the magnetopause oscillations
were found to be azimuthally stationary. This range is stable in time
for the duration of the oscillations. The results therefore demonstrate
that the stationary nature of MSE azimuthally is due to a balance
of the surface wave Poynting flux directed towards the subsolar point
opposing the tailward magnetosheath flow. Outside of this region,
however, even when the Poynting flux is in opposition to the magnetosheath
flow it is unable to overcome advection and thus travelling surface
waves result.

\subsection*{Analytic MHD theory}

\begin{figure}
\begin{centering}
\includegraphics[scale=0.9]{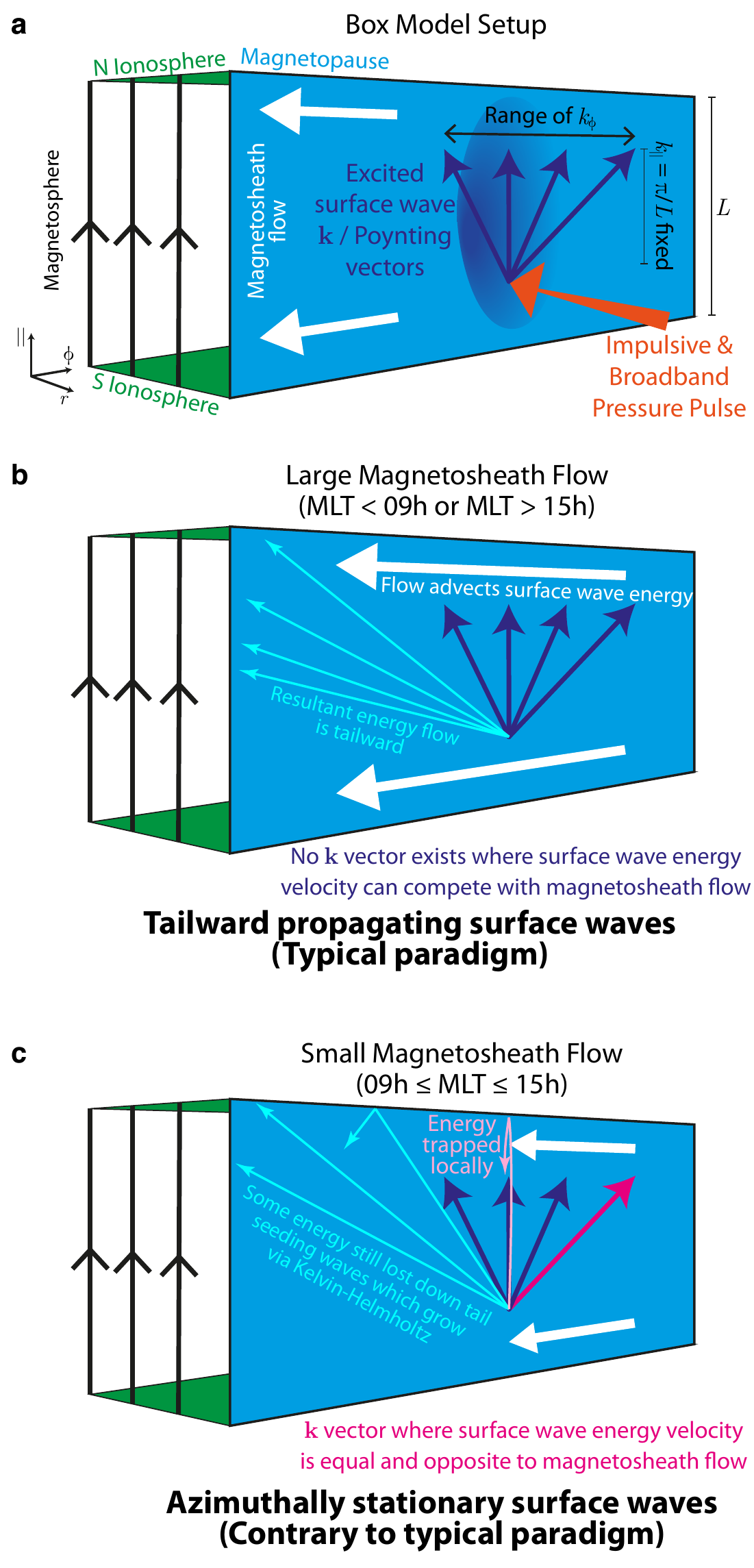}
\par\end{centering}
\caption{\textbf{Cartoon illustrating the results of this study}. Panel~a
shows the box model magnetosphere, magnetosheath flow (white), and
surface mode wavevectors (dark blue) excited by the pressure pulse
(orange). Subsequent panels depict the resultant energy flow (lighter
coloured arrows) of the surface mode wavevectors for b) large and
c) small magnetosheath flows.\label{fig:cartoon}}
\end{figure}

Finally, we look to incompressible MHD theory (where $k_{r}^{2}+k_{\phi}^{2}+k_{\parallel}^{2}=0$
for surface waves \cite{pu83,plaschke11}) to understand this picture
of the energy flow and azimuthal propagation present within MSE. We
consider a fundamental mode magnetopause surface wave of amplitude
$A$ in displacement within a box model magnetosphere with homogeneous
half-spaces as depicted in Figure~\ref{fig:cartoon}a. The azimuthal
component of the Poynting vector at the equator on the magnetosphere
side of the boundary for northward IMF is given by (equation~15 of
\cite{juninger85})
\begin{equation}
\left\langle S_{\phi,sph}\right\rangle =A^{2}\omega k_{\phi}\frac{k_{\parallel}^{2}}{k_{\phi}^{2}+k_{\parallel}^{2}}\frac{B_{0,sph}^{2}}{2\mu_{0}}\exp\left(-2\left|\mathfrak{Im}\left(k_{r}\right)\right|\left|r-r_{mp}\right|\right)
\end{equation}
with its equivalent on the magnetosheath side being $B_{0,msh}^{2}/B_{0,sph}^{2}$
times this and thus negligible. The wave energy densities are (following
equations~11 and 13 of \cite{juninger85})
\begin{equation}
\begin{array}{ccccc}
\left\langle u_{sph}\right\rangle  & \approx & \left\langle u_{B,sph}\right\rangle  & = & \frac{B_{0,sph}^{2}}{4\mu_{0}}A^{2}\frac{k_{\parallel}^{4}}{k_{\phi}^{2}+k_{\parallel}^{2}}\exp\left(-2\left|\mathfrak{Im}\left(k_{r}\right)\right|\left|r-r_{mp}\right|\right)\\
\left\langle u_{msh}\right\rangle  & \approx & \left\langle u_{K,msh}\right\rangle  & = & \frac{1}{4}\rho_{0,msh}\omega^{2}A^{2}\frac{2k_{\phi}^{2}+k_{\parallel}^{2}}{k_{\phi}^{2}+k_{\parallel}^{2}}\exp\left(-2\left|\mathfrak{Im}\left(k_{r}\right)\right|\left|r-r_{mp}\right|\right)
\end{array}
\end{equation}
Constructing the net energy velocity of the surface wave and simplifying
using equation~\ref{eq:mse-frequency} gives

\begin{align}
\mathbf{v}_{E,tot} & =\frac{\left\langle \mathbf{S}_{sph}\right\rangle +\left\langle \mathbf{S}_{msh}\right\rangle +\left\langle u_{msh}\right\rangle \mathbf{v}_{0,msh}}{\left\langle u_{sph}+u_{msh}\right\rangle }\label{eq:vE1}\\
 & \approx\frac{\omega k_{\phi}k_{\parallel}^{2}\frac{B_{0,sph}^{2}}{2\mu_{0}}+\frac{1}{4}\rho_{0,msh}\frac{B_{0,sph}^{2}}{\mu_{0}\rho_{0,msh}}k_{\parallel}^{2}\left(2k_{\phi}^{2}+k_{\parallel}^{2}\right)v_{0,msh}}{\frac{B_{0,sph}^{2}}{4\mu_{0}}\left[k_{\parallel}^{4}+\omega^{2}\frac{k_{\parallel}^{2}}{\omega^{2}}\left(2k_{\phi}^{2}+k_{\parallel}^{2}\right)\right]}\hat{\boldsymbol{\phi}}\label{eq:vE2}\\
 & \approx\left[\frac{\omega k_{\phi}}{k_{\phi}^{2}+k_{\parallel}^{2}}+\frac{2k_{\phi}^{2}+k_{\parallel}^{2}}{2\left(k_{\phi}^{2}+k_{\parallel}^{2}\right)}v_{0,msh}\right]\hat{\boldsymbol{\phi}}\label{eq:vE3}
\end{align}
By setting this to zero, i.e. no net azimuthal energy flow, and solving
for real azimuthal wavenumbers yields the requirement
\begin{equation}
\frac{\omega}{k_{\parallel}}\geq\sqrt{2}v_{0,msh}
\end{equation}
This sets a limit on where it is possible for surface wave energy
to be trapped locally due to the speed of the adjacent magnetosheath.
We can frame this limit purely in terms of solar wind and magnetosheath
conditions using equation~\ref{eq:mse-frequency-sw} as
\begin{equation}
v_{0,msh}\leq\sqrt{\frac{\rho_{sw}}{\rho_{msh}}}v_{sw}
\end{equation}
According to gas-dynamic models of magnetosheath plama conditions
\cite{spreiter66} this is satisfied for 08:40--15:20h~MLT, in excellent
agreement with the stationary region in the global MHD simulation.
This extent should vary only slightly with solar wind conditions (based
on previous magnetosheath and MSE variability studies \cite{archer15,walsh12}),
however, future work could test this.

\section*{Discussion}

In this paper we show that the recently discovered magnetopause surface
eigenmode (MSE), the lowest frequency normal mode of a magnetosphere,
does not conform to the well-established paradigm in global magnetospheric
dynamics of tailward propagation. Multi-spacecraft observations, global
MHD simulations, and analytic MHD theory are employed. Both the observations
and simulation show Poynting vectors in the magnetosphere which point
towards the subsolar point across the dayside, contrary to current
models of the magnetospheric response to impulsive driving \cite{sibeck90}.
This energy flux thus opposes advection by the magnetosheath and we
find from the simulation that these two cancel one another in the
region 09--15h magnetic local time, resulting in an azimuthally stationary
surface wave. Outside of this region, however, the waves travel tailward.
Considering surface wave energy fluxes in a simple box model of the
magnetosphere shows excellent agreement with the simulation on the
conditions required for a stationary wave to be possible. Our conclusions
are summarised in Figure~\ref{fig:cartoon} within this box model.
When an impulsive solar wind transient arrives at the magnetopause,
its broadband nature excites surface waves on the boundary with a
wide range of frequencies $\omega$ and wave vectors $\mathbf{k}$.
The boundary conditions at the northern and southern ionospheres quantise
the possible values of $k_{\parallel}$ largely determining $\omega$,
however, $k_{\phi}$ will be unconstrained as depicted in panel~a.
For large magnetosheath flow speeds (panel~b) none of the excited
wave vectors are able to compete with advection and the resultant
motion is tailward, in line with expectations. In the regime of small
magnetosheath flow speeds (panel~c), however, there exists an excited
$k_{\phi}$ in opposition to the magnetosheath flow which is able
to exactly balance its advective effect. This leads to surface wave
energy being trapped locally as an azimuthally stationary wave. All
waves of other $k_{\phi}$ will be lost down the tail. This picture
not only explains the global propagation of magnetopause surface waves
but also how MSE on the dayside can seed fluctuations into the magnetospheric
flanks. The simulation shows that these seeded waves which originate
on the dayside subsequently grow in amplitude via the Kelvin-Helmholtz
instability, despite being at a lower frequency to its intrinsic one,
and may couple to cavity/waveguide and Alfv\'{e}n modes in regions
of the magnetosphere where their frequencies match. This reveals MSE's
effects are not confined merely to the dayside (standing) region,
instead having global effects on the magnetosphere as its most fundamental
normal mode.

The cartoon highlights that, at each location on the boundary, after
the other (blue) wavevectors have been swept downtail and the boundary
has formed its resonance, the physics of the azimuthally stationary
surface wave is confined to a small local time region, i.e. a single
meridian of geomagnetic field lines. While the initial perturbation
on the boundary and the corresponding transient response will depend
on the specifics of the driving pressure variation (scale sizes, location
of impact etc.), one can simply decompose the initial perturbation
at each local time into the normal modes along the field (MSE) and
this should entirely dictate the subsequent resonant response at that
local time. Indeed, the local boundary motion and Poynting fluxes
were in agreement across both observations and simulations despite
different scale size drivers being leveraged. The locality to the
physics means that the azimuthally stationary wave should be limited
to the local times in which the driver impacted the magnetopause,
hence the scale of the driver in azimuth (within the 09--15h local
time range) would be imprinted in the stationary waves excited.

These results raise the question why only tailward propagating dynamics
are reported in current models and observations of the magnetospheric
response to impulsive driving \cite{sibeck90}. It is clear that the
models do not incorporate the possibility of surface wave reflection
due to bounding by the ionosphere, which is key to our results, since
while this was proposed long ago \cite{chen74} it has only recently
been discovered \cite{archer19,He2020}. MSE constitute the lowest
possible frequency normal mode of the magnetospheric system and its
fundamental frequencies can often be fractions of milliHertz \cite{archer15}.
Such long period narrowband waves are challenging to identify observationally
in general, either by orbiting spacecraft or ground-based measurements,
due to potential spatio-temporal mixing \cite{urban16} and the difficultly
in distinguishing from turbulence/noise \cite{dimatteo17,dimatteo18}.
For these reasons global magnetospheric dynamics and their associated
ULF waves have often concentrated on the continuous pulsation (Pc)
bands above $2\,\mathrm{mHz}$ \cite{jacobs64}, which would not typically
incorporate the effects presented here. Furthermore, Figure~\ref{fig:cartoon}
shows that the majority of surface wave energy excited by the driver
does still propagate tailward, with only a small amount being trapped
locally that propagate against the flow. Therefore, if further upstream
driving by pressure variations occurs during these oscillations, the
superposition of waves present could easily mask the sunward Poynting
vectors associated with MSE thereby showing only a net tailward energy
flow. Future work is required in developing less restrictive observational
criteria for the detection of MSE in general and to undertake statistical
studies of MSE occurrence to better understand how common this mode,
and the results presented on its energy flow, may be in reality to
the variety of impulsive drivers that impact on geospace.

The global waves associated with this normal mode of Earth's magnetosphere,
possible due to the surface wave propagation against the magnetosheath
flow, will have important implications upon radiation belt dynamics
\cite{elkington06,summers13}. The large-scale oscillations of the
magnetopause may cause the further shadowing of radiation belt electrons
than predicted simply by a pressure balanced quasi-static response
to the driver. Furthermore, MSE's ULF wave signatures present coherent
and slowly varying perturbations in compressional magnetic fields
and azimuthal electric fields which deeply penetrate across the entire
dayside magnetosphere, which may be ideal for the drift-resonant interaction
and/or radial diffusion of radiation belt particles. However, current
methods of understanding these processes are suited only to the inner
magnetosphere since they assume azimuthal symmetry, therefore, more
work is required in assessing the impact on the radiation belts of
this normal mode and asymmetric outer magnetospheric waves in general.
While the observations confirm significant energy flow along the magnetic
field towards the polar regions, the wide region of stationarity from
the simulations suggests weak coupling of the surface waves to the
Alfv\'{e}n mode across the dayside. This implies MSE have auroral,
ionospheric, and ground magnetometer signatures unlike other known
ULF waves, with these remaining poorly understood \cite{archer19,He2020}.
In the flanks, however, it seems likely that MSE-seeded waves could
at times easily be mistaken for intrinsic Kelvin-Helmholtz waves or
waveguide modes, despite the origin of the fluctuations on the dayside
as shown in the simulation. These factors may be why previous ground-based
searches, through widely-used diagnostics for other wave modes, called
MSE's existence into question \cite{pilipenko17,pilipenko18}. The
work thus highlights that care needs to be taken in understanding
the mechanisms which result in various dynamical modes in near-Earth
space since they can all be intimately coupled.

Surface waves are known to be present at the other planetary magnetospheres
\cite{sundberg12,delamere16}, which span a vast range of sizes, morphologies,
and plasma conditions \cite{bagenal13}. The surface eigenmode in
principle should be a universal feature of boundaries in magnetospheres
\cite{chen74}, and thus the simple analytic theory presented here
(in the magnetospheric reference frame) may be instructive in assessing
where and in what frequency ranges these fundamental dynamics of the
boundary may be prevalent at other environments. The simple predictions
could then be compared to tailored global MHD simulations of these
systems as well as spacecraft observations.

Many other space and astrophysical systems too exhibit surface waves
where, like in the case of a magnetopause, substantial background
flows may be present. A notable example are the sausage and kink modes
of coronal loops, which share many conceptual similarities to the
surface eigenmodes -- they are standing (though sometimes propagating)
transverse oscillations of the dense flux tubes in coronal active
regions, anchored on both ends by the chromosphere, excited by loop
displacements from coronal eruptions or shear flows in coronal plasma
non-uniformities \cite{li13,yu20}. Asymmetric and/or inhomogeneneous
flows around/along these structures affect surface wave evolution,
with important space weather consequences such as causing coronal
mass ejections to turn away from their original propagation direction
\cite{foullon13}, though these effects are not typically incorporated
into models of coronal loop oscillations. Our results from \textit{in
situ} observations at the magnetopause (not possible for the corona
and other space/astrophysical environments) challenge the paradigm
that surface waves necessarily propagate in the direction of the driving
flow/pressure, as when discontinuities are bounded the trapping of
surface waves may occur in opposition to advective effects, allowing
these waves to form across broader regions and to persist longer than
would otherwise be expected. The work may therefore have insights
into the structure and stability of these universal dynamical modes
elsewhere.

\section*{Methods}

\subsection*{Poynting's theorem for MHD waves}

Energy conservation for MHD wave perturbations (denoted by $\delta$'s
with subscript $0$'s representing equilibrium values) involves the
wave energy density
\begin{equation}
u=u_{B}+u_{K}=\frac{\left|\delta\mathbf{B}\right|^{2}}{2\mu_{0}}+\frac{1}{2}\rho_{0}\left|\delta\mathbf{v}\right|^{2}\label{eq:energy-density}
\end{equation}
(consisting of magnetic $u_{B}$ and kinetic $u_{K}$ contributions)
and wave energy flux given by the Poynting vector 
\begin{equation}
\mathbf{S}=\frac{\delta\mathbf{E}\times\delta\mathbf{B}}{\mu_{0}}\label{eq:Poynting}
\end{equation}
\cite{walker05} where $\mathbf{E}$ is the electric field. Time-averaging
and taking their ratio yields the so-called energy velocity
\begin{equation}
\mathbf{v}_{E}=\frac{\left\langle \mathbf{S}\right\rangle }{\left\langle u\right\rangle }\label{eq:energy-velocity}
\end{equation}
, equivalent to the group velocity for stable waves \cite{bers00}.
In a moving medium, wave energy advects with the background plasma,
giving an additional flux $u\mathbf{v}_{0}$. These principles are
applied throughout.

\subsection*{Spacecraft observations}

Observations in this paper are taken from the Time History of Events
and Macroscale Interactions (THEMIS; \cite{angelopoulos08}) spacecraft
taken during the previously reported interval of MSE \cite{archer19}.
The five spacecraft were close to the equilibrium magnetopause in
a string-of-pearls formation. Data from the fluxgate magnetometer
(FGM) \cite{auster08}, electrostatic analyser (ESA) \cite{mcfadden08a},
and electric field (EFI) \cite{bonnell08} instruments are used. Note
for the latter we use the $\mathbf{E}\cdot\mathbf{B}=0$ approximation
(valid over ULF timescales) for THD and THE to replace the measured
axial fields at each time, however the instrument was not yet deployed
by THA so $\mathbf{E}=-\mathbf{v}\times\mathbf{B}_{0}$ is used, which
was found to be reliable for the other spacecraft. We note that THA
plasma measurements were not available prior to 22:08~UT. Magnetosheath
intervals have been removed from THA, THD and THE observations, identified
when the electron density was greater than $5\,\mathrm{cm}^{-3}$
or the magnetic field strength was less than $45\,\mathrm{nT}$. Vectors
within the magnetosphere have been rotated into local orthogonal field-aligned
coordinates ($r,\phi,\bigparallel$). The field-aligned direction
($\parallel$) is given based on a robust linear regression of the
magnetic field vectors \cite{huber81,street88}, with the azimuthal
($\phi$) direction being the cross product of $\parallel$ with the
spacecraft's geocentric position thus pointing eastward, and the radial
($r$) direction completing the right-handed set directed away from
the Earth. While this coordinate rotation may result in some small
$E_{\parallel}$, these are negligible compared to the other components
and do not influence the results.

\subsection*{Global MHD simulations}

We reproduce a high-resolution ($\nicefrac{1}{8}\text{\textendash}\nicefrac{1}{16}\,\mathrm{R_{E}}$
in the regions considered in this paper, see Supplementary~Figure~4
for grid) Space Weather Modeling Framework (SWMF; \cite{toth05,toth12})
simulation run of MSE excited by a $1\,\mathrm{min}$ solar wind density
pulse (with sunward normal) under northward IMF \cite{hartinger15}.
Full details of the run are given in Supplementary~Table~1. For
all simulation quantities, perturbations are defined as the difference
to the linear trend before ($t=0\,\mathrm{min}$) and after ($t=60\,\mathrm{min}$)
the response to the pulse. Vectors are rotated into similar local
field-aligned coordinates. The magnetopause location is determined
as the last closed field line along geocentric rays through a bisection
method accurate to $0.01\,\mathrm{R_{E}}$. The bow shock standoff
distance has been identified via interpolation as the point where
the density is half that in the solar wind. In displaying perturbations
in the simulation, a bi-symmetric log transform \cite{webber12} is
often employed due to the much larger amplitudes present during compression
and rebound phases.

\subsection*{Time-based filtering}

A time-based filtering technique is used to extract MSE wave perturbations
and suppress noise and higher/lower frequency signals. This was chosen
to avoid the potential influence of ringing artefacts or edge effects
when using frequency-based methods due to the nonstationary process.
Nonetheless, several different filtering methods were tested and the
main results of the paper remained robust.

In the method presented for the spacecraft observations, first the
raw data is smoothed using a $400\,\mathrm{s}$ robust LOESS method
\cite{cleveland79}. For stationary processes this has a corresponding
cutoff frequency of $3.6\,\mathrm{mHz}$, therefore retains both the
$1.8\,\mathrm{mHz}$ fundamental and $3.3\,\mathrm{mHz}$ second harmonic
MSE signals present \cite{archer19}. To remove any lower frequency
trends still present, the mean envelope from cubic Hermite interpolation
\cite{fritsch80} is subtracted. These effectively bandpass filtered
quantities are used for calculating the instantaneous wave Poynting
vectors and energy densities. A time-averaging method is performed
also by using the mean envelope from interpolation. The time-based
methods used also allow uncertainties to be estimated. This is done
via a running root-mean-squared (RMS) deviation between the raw and
LOESS smoothed time-series, which are then propagated through the
subsequent methods used \cite{horst81,white17}.

To extract the MSE signal from the simulation, either in magnetopause
location or grid point data, we firstly neglect the initial large
amplitude compression and rebound. This is done by only using data
from half an MSE period after the magnetopause's return to equilibrium,
i.e. after the dotted line in Figure~\ref{fig:Magnetopause-motion}a.
For grid point data the timing at the magnetopause with the same X
coordinate is used. The secondary spectral peak is then suppressed
using the same filtering procedure as for the THEMIS data. The only
differences are that standard (rather than robust) LOESS is used due
to reduced temporal resolution, and the window size used was $570\,\mathrm{s}$
corresponding to a $2.4\,\mathrm{mHz}$ cutoff.

\subsection*{Fourier and wavelet techniques}

To compute time-averaged Poynting vectors as a function of frequency
a standard complex Fourier approach is used (equation 2 of \cite{hartinger13})
\begin{equation}
\left\langle \mathbf{S}\left(\omega\right)\right\rangle =\frac{\mathrm{\mathfrak{Re}}\left(\mathbf{E}\left(\omega\right)\times\mathbf{B}^{*}\left(\omega\right)\right)}{2\mu_{0}}
\end{equation}
This is done both in frequency-space using Welch's method \cite{welch67}
in computing one-sided cross spectra, i.e. the products of electric
and magnetic field components, and in time-frequency space using products
of analytic Morse continuous wavelet transforms \cite{lilly12}. In
both cases, a null hypothesis of autoregressive noise is assumed,
where the AR(1) parameters for each component of the electric and
magnetic fields are estimated using constrained maximum likelihood
and 500 independent Monte Carlo simulations are performed based on
these models \cite{box}, with 95\% confidence intervals being constructed
by taking percentiles (2.5\% and 97.5\%) of the resulting time-averaged
Poynting vectors.

\subsection*{Slowness}

To quantify the propagation of MSE boundary perturbations, the slowness
was computed through cross correlating the filtered magnetopause signals
between adjacent local times. By interpolating the peak to find its
corresponding time lag $\Delta t$, the azimuthal slowness is given
by

\begin{equation}
s_{\phi}=\frac{\Delta t}{\left|\Delta\mathbf{r}\right|}
\end{equation}
where $\left|\Delta\mathbf{r}\right|$ is the distance between the
two points on the boundary used. Standard errors in the correlation
coefficient were also calculated and propagated through the interpolation
procedure to arrive at uncertainties.

\section*{Data availability}

The THEMIS spacecraft data are available at \url{http://themis.ssl.berkeley.edu/data/themis/}
where level-2 data from the FGM, ESA, and EFI instruments on each
spacecraft has been used in this study. The SWMF simulation data generated
in this study are available in the Community Coordinated Modeling
Center (CCMC) at \url{https://ccmc.gsfc.nasa.gov/results/viewrun.php?domain=GM&runnumber=Michael_Hartinger_061418_1}.

\section*{Code availability}

The SWMF and BATS-R-US (Block-Adaptive Tree Solarwind Roe-type Upwind
Scheme) software is available at \url{https://github.com/MSTEM-QUDA}.
The SWMF and BATS-R-US tools used are available at \url{https://ccmc.gsfc.nasa.gov}.

\bibliographystyle{unsrt}
\bibliography{msestructure}

\section*{Acknowledgements}

MOA holds a UKRI (STFC / EPSRC) Stephen Hawking Fellowship EP/T01735X/1.
MDH was supported by NASA grant 80NSSC19K0127. FP was supported by
the Austrian Science Fund (FWF): P 33285-N. DJS was supported by STFC
grant ST/S000364/1. We acknowledge NASA contract NAS5-02099 for use
of data from the THEMIS Mission. Specifically K.~H. Glassmeier, U.
Auster and W. Baumjohann for the use of FGM data provided under the
lead of the Technical University of Braunschweig and with financial
support through the German Ministry for Economy and Technology and
the German Center for Aviation and Space (DLR) under contract 50~OC~0302;
C.~W. Carlson and J.~P. McFadden for use of ESA data; and J.~W.
Bonnell and F.~S. Mozer for EFI data. Simulation results have been
provided by the Community Coordinated Modeling Center (CCMC) at Goddard
Space Flight Center using the SWMF and BATS-R-US tools developed at
the University of Michigan's Center for Space Environment Modeling
(CSEM).

\section*{Author contributions}

MOA and MDH conceived the study. LR gave technical support. MOA performed
the analysis of the data. MOA, MDH, FP, and DJS interpreted the results.
MOA wrote the paper, with MDH, FP, and DJS assisting.

\section*{Competing interests statement}

The authors declare no competing interests.
\end{document}

\title{Supplementary information for ``Magnetopause ripples going against
the flow form azimuthally stationary surface waves''}
\author{M.~O.~Archer, M.~D.~Hartinger, F.~Plaschke, D.~J.~Southwood,
\& L. Rastaetter}

\maketitle
\renewcommand{\figurename}{Supplementary Fig.} 

\renewcommand{\tablename}{Supplementary Table} \clearpage{}

\begin{table}[H]
\begin{centering}
\begin{tabular}{lcc}
\multicolumn{3}{l}{Simulation Setup}\tabularnewline
\multicolumn{2}{l}{Dipole GSM Orientation} & (0,0,1)\tabularnewline
\multicolumn{2}{l}{Dipole Update} & No\tabularnewline
\noalign{\vskip6pt}
\multicolumn{3}{l}{Solar Wind Conditions}\tabularnewline
Quantity & Solar Wind & 1~min Density Pulse\tabularnewline
$n$ ($\mathrm{cm}^{-3}$) & 6 & 14\tabularnewline
$T$ ($\mathrm{K}$) & 116,174.0 & 49,788.8\tabularnewline
$\mathbf{v}_{GSM}$ ($\mathrm{km}\,\mathrm{s}^{-1}$) & (-450,0,0) & (-450,0,0)\tabularnewline
$\mathbf{B}_{GSM}$ ($\mathrm{nT}$) & (0,0,5) & (0,0,5)\tabularnewline
\multicolumn{2}{l}{GSM normal} & (1,0,0)\tabularnewline
\noalign{\vskip6pt}
\multicolumn{3}{l}{Ionospheric Conditions}\tabularnewline
\multicolumn{2}{l}{Conductivity} & 5~mho (uniform)\tabularnewline[10pt]
\end{tabular}
\par\end{centering}
\caption{\textbf{Details of the SWMF (Space Weather Modeling Framework) global
magnetohydrodynamic simulation run used in this paper. }The first
subgroup details the setup of the magnetic dipole in Geocentric Solar
Magnetospheric (GSM) coordinates. The second subgroup lists the plasma
number density $n$, temperature $T$, velocity $\mathbf{v}$, and
magnetic field $\mathbf{B}$ in both the ambient solar wind and the
1~min density pulse, along with the latter's orientation. The final
subgroup highlights the ionospheric conditions used. \label{tab:run-details}}
\end{table}

\begin{figure*}
\begin{centering}
\noindent \makebox[\textwidth]{\raisebox{0pt}[21cm]{\includegraphics[width=0.975\paperwidth]{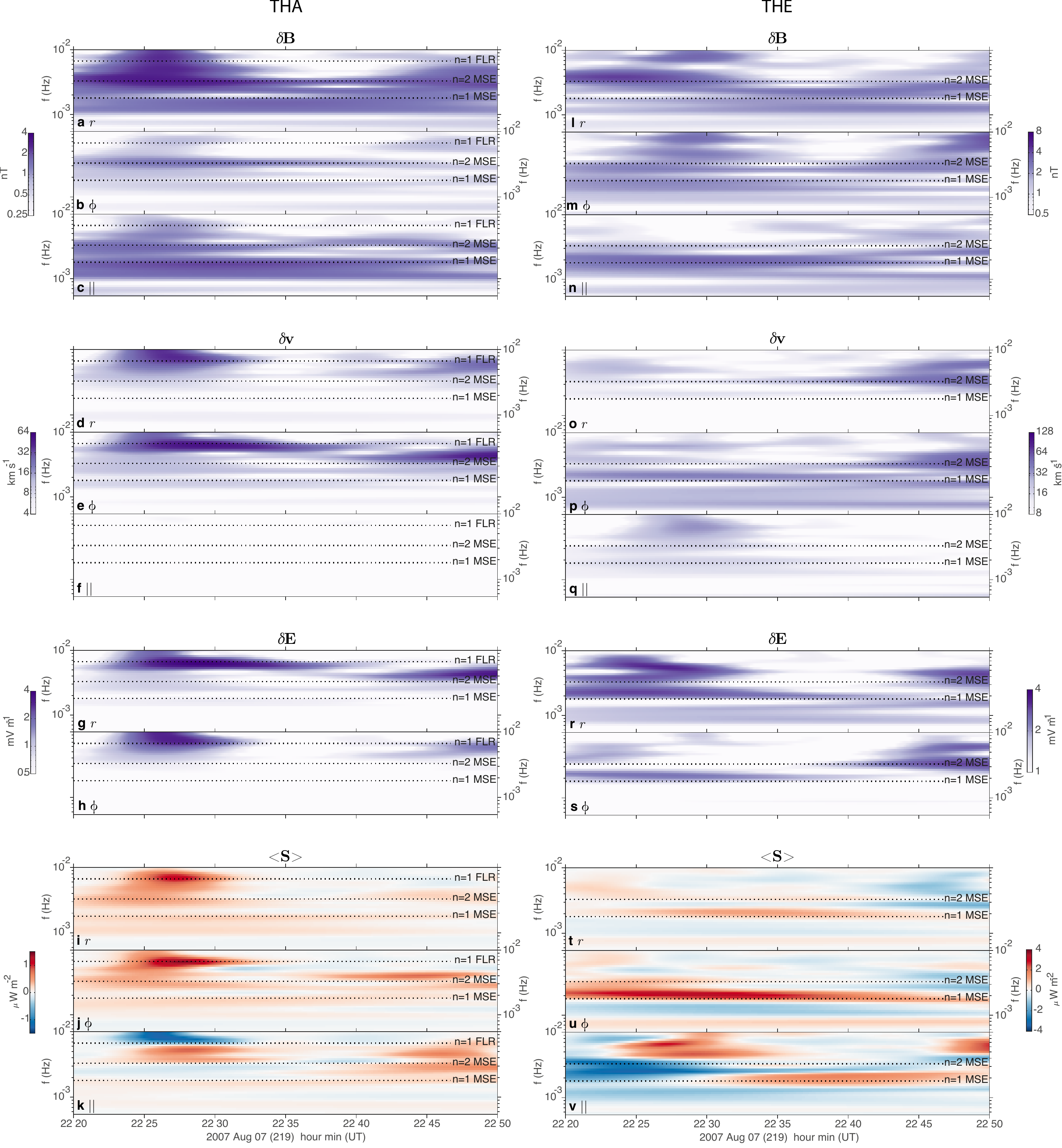}}}
\par\end{centering}
\caption{\textbf{Dynamic spectra of the spacecraft observations using the wavelet
transform}. Shown for THA (a--k) and THE (l--v). Panels show radial
(top), azimuthal (middle), and field-aligned (bottom) components of
the magnetic (a--c, l--n), velocity (d--f, o--q), and electric
(g--h, r--s) field perturbations, with the logarithmic colour scales
indicating amplitudes. The components of the time-averaged Poynting
vector as function of frequency and time are depicted in the final
panels (i--k, t--v). Dotted lines depict MSE and local FLR frequencies.}

\end{figure*}

\begin{figure*}
\begin{centering}
\noindent \makebox[\textwidth]{\raisebox{0pt}[20cm]{\includegraphics[scale=1.1]{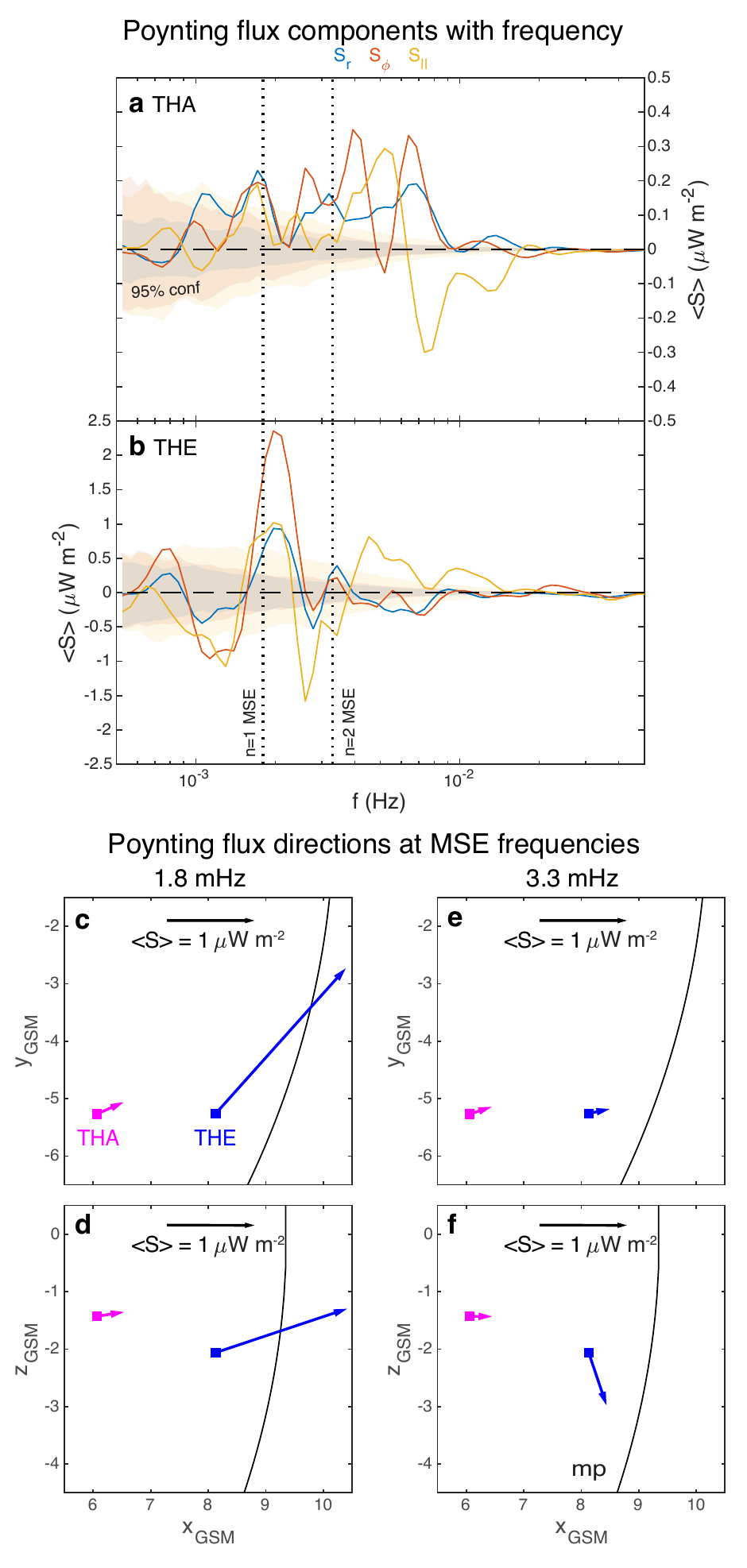}}}\caption{\textbf{Poynting fluxes averaged over the interval}. These are shown
by component as a function of frequency for THA (a) and THE (b). Shaded
areas indicate 95\% confidence intervals for null hypotheses of autoregressive
noise for each component. The two MSE frequencies are highlighted
by vertical dotted lines. Subsequent panels show, at the two MSE frequencies,
the Poynting fluxes in the $z_{GSM}=-2.1\,\mathrm{R_{E}}$ (c,e) and
$y_{GSM}=-5.3\,\mathrm{R_{E}}$ (d,f) planes as arrows originating
from the spacecraft locations (squares). A model magnetopause is also
shown (black).}
\par\end{centering}
\end{figure*}

\begin{figure}
\begin{centering}
\noindent \makebox[\textwidth]{\includegraphics[scale=1.5]{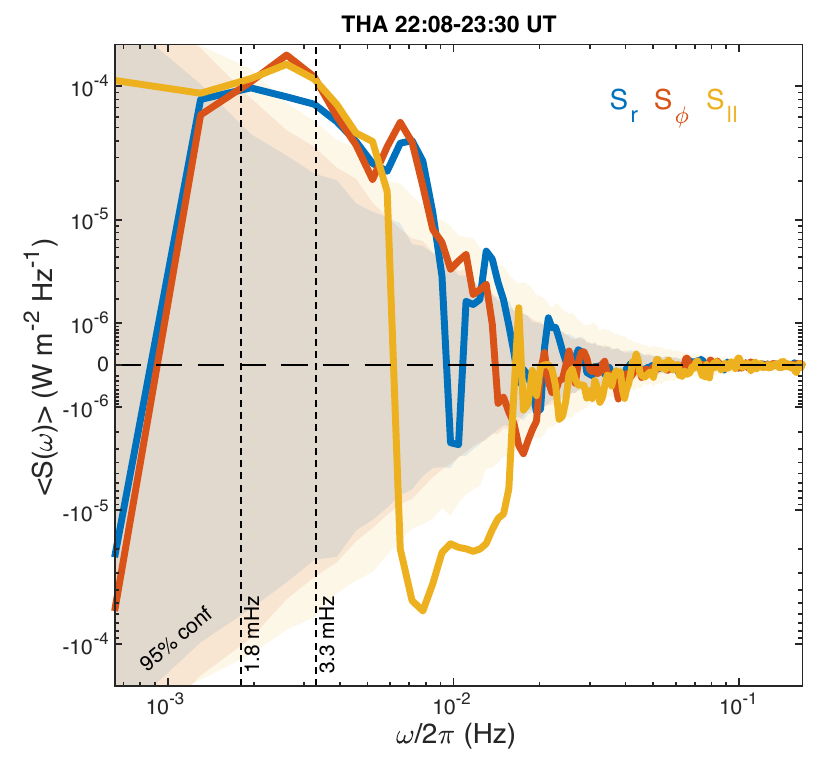}}
\par\end{centering}
\caption{\textbf{Components of the time-averaged Poynting vector at THA as
a function of frequency over an extended time interval}. A bi-symmetric
log scale is used on the vertical axis. Shaded coloured areasindicate
95\% confidence intervals for null hypotheses of autoregressive noise
for each component. The identified MSE frequencies are indicated by
vertical dotted lines.}

\end{figure}

\begin{figure}
\begin{centering}
\noindent \makebox[\textwidth]{\includegraphics[scale=1.5]{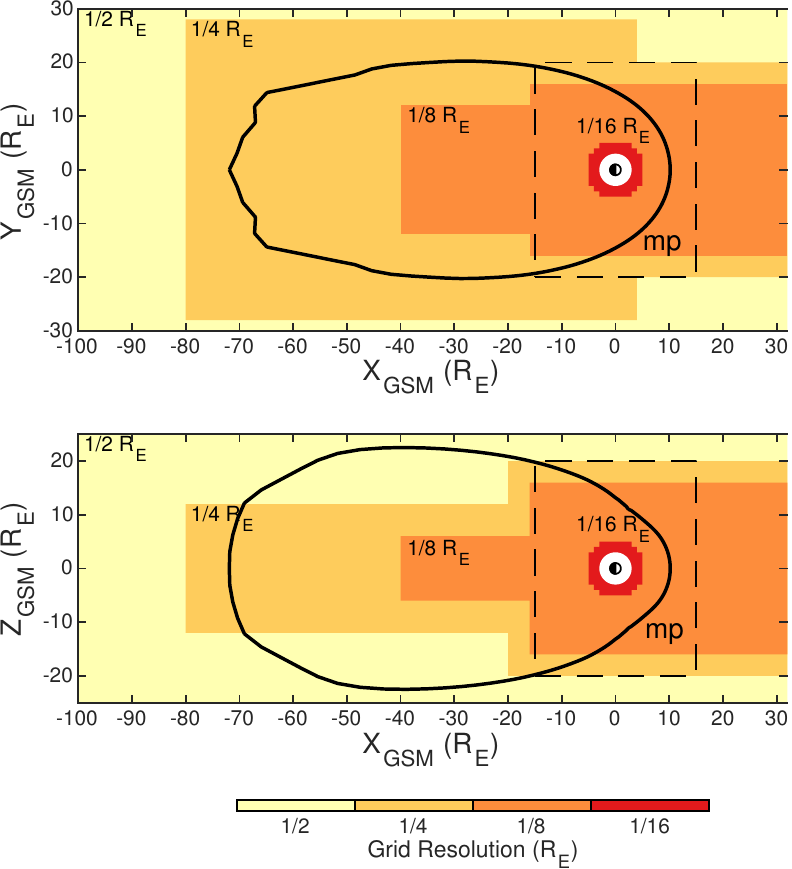}}\caption{\textbf{Grid resolution of the global MHD simulation}. Shown for the
GSM XY (top) and XZ (bottom) planes. The black line depicts the equilibrium
boundary of closed field lines, used as a proxy for the magnetopause
location in this paper. The dashed box indicates the region of interest
in this paper.}
\par\end{centering}
\end{figure}